\documentclass[a4paper,twocolumn,11pt,accepted=2017-05-09]{quantumarticle}
\pdfoutput=1
\usepackage[utf8]{inputenc}
\usepackage[english]{babel}
\usepackage[T1]{fontenc}
\usepackage{amsmath}
\usepackage{hyperref}

\usepackage{graphicx}
\usepackage{amsmath}
\usepackage{tikz}
\usetikzlibrary{quantikz}
\usepackage{physics}
\usepackage{lipsum}
\begin{document}

\title{PCA and t-SNE analysis in the study of QAOA entangled and non-entangled mixing operators}

\author{Brian García Sarmina}
\affiliation{CIC, Instituto Politécnico Nacional,  07738 Ciudad de México, México}
\orcid{0000-0002-0182-7312}
\email{brian.garsar.6@gmail.com}

\author{Guo-Hua Sun}
\affiliation{CIC, Instituto Politécnico Nacional,  07738 Ciudad de México, México}
\orcid{0000-0002-0689-2754}
\email{gsun@cic.ipn.mx}

\author{Shi-Hai Dong}
\affiliation{CIC, Instituto Politécnico Nacional,  07738 Ciudad de México, México}
\orcid{0000-0002-0769-635X}
\email{dongsh2@yahoo.com}
\maketitle

\begin{abstract}
In this paper, we employ PCA and t-SNE analysis to gain deeper insights into the behavior of entangled and non-entangled mixing operators within the Quantum Approximate Optimization Algorithm (QAOA) at varying depths. Our study utilizes a dataset of parameters generated for max-cut problems using the Stochastic Hill Climbing with Random Restarts optimization method in QAOA. Specifically, we examine the $RZ$, $RX$, and $RY$ parameters within QAOA models at depths of $1L$, $2L$, and $3L$, both with and without an entanglement stage inside the mixing operator. The results reveal distinct behaviors when we process the final parameters of each set of experiments with PCA and t-SNE, where in particular, entangled QAOA models with $2L$ and $3L$ present an increase in the amount of information that can be preserved in the mapping. Furthermore, certain entangled QAOA graphs exhibit clustering effects in both PCA and t-SNE. Overall, the mapping results clearly demonstrate a discernible difference between entangled and non-entangled models, quantified numerically through explained variance in PCA and Kullback-Leibler divergence (after optimization) in t-SNE, where some of these differences are also visually evident in the mapping data produced by both methods.

\end{abstract}

\section{Introduction}

The analysis of Variational Quantum Algorithms (VQA), including the Quantum Approximate Optimization Algorithm (QAOA), has received significant attention in recent years \cite{R1, R2, R3, R4, R27}. Understanding how these algorithms explore problem spaces and the relationships between rotation gates in the circuit and algorithm performance is an exciting area of research. Previous studies have provided insights into these aspects \cite{R15, R16}, as well as exploring the connection between problem Hamiltonian structures and generated landscapes \cite{R28, R29}. However, as quantum hardware advances and circuits become more complex, a deeper understanding of these algorithms becomes increasingly crucial \cite{R24, R26}.

The analysis of QAOA is influenced by factors such as circuit depth and optimization strategies, which can impact result accuracy. Research in this area often focuses on problem landscape representation, crucial for understanding QAOA from a problem-solving perspective \cite{R2, R4, R17, R25}. However, it is essential to consider other aspects when studying QAOA, including extracting information about underlying models, their limitations, and strengths, before applying them to specific problems.

Therefore, this paper aims to contribute to existing knowledge by analyzing the behavior of entangled and non-entangled mixing operators in QAOA. We utilize PCA and t-SNE techniques to study QAOA models at different depths (1L, 2L, and 3L) applied to specific max-cut problems. Our objective is to identify distinct behaviors that offer insights into how QAOA responds to different problems and visualize these behaviors practically. Notable examples of visualization studies in VQAs include the works of \textit{Moussa et al. (2022)} \cite{R18} and \textit{Rudolph et al. (2021)} \cite{R29}. In \textit{Moussa et al. (2022)}, t-SNE was used as a preprocessing step to reveal clustering tendencies in QUBO problems and aid in determining QAOA parameters. They also explored supervised techniques when clusters did not adequately represent corresponding points, resulting in improved prediction of QAOA parameters. \textit{Rudolph et al. (2021)} performed an analysis of various visualization techniques, including PCA, applied to different VQAs. Their focus was on generating mappings of the optimization landscape for specific problems, studying parameter concentration in QAOA, and investigating other phenomena.


In our study, our primary focus was on the representation, visualization, and information extraction of the resulting QAOA parameters (after optimization) using PCA and t-SNE. Our aim was to evaluate the effectiveness of these visualization strategies in providing comprehensive insights into the underlying models. We explored both graphical representations and internal metrics derived from PCA and t-SNE techniques to assess the information they can offer. By analyzing the visualizations and examining the derived metrics, we aimed to determine whether these strategies can provide sufficient information about the QAOA models being studied.

\section{PCA and t-SNE description}

In this section, we describe how the PCA and t-SNE approach is employed to analyze the properties of the various QAOA models.

\subsection{PCA}

Principal Component Analysis (PCA) is a dimensionality reduction method that aims to capture the maximum amount of information from a dataset while reducing its dimensionality \cite{R19, R20}. It achieves this by identifying the principal components, which are specific structures that capture the most variance in the dataset. The principal components are computed by finding the eigenvectors of the covariance matrix of the dataset. These eigenvectors represent the directions in which the data varies the most. The corresponding eigenvalues indicate the amount of variance captured along each principal component.
By transforming the original data into the new coordinate system defined by the principal components, we can visualize the data in a more meaningful way. Typically, the first two principal components are used for visualization purposes, as they capture the most significant variance in the dataset. This transformation allows us to represent the data in a lower-dimensional space, while still preserving as much of the original information as possible \cite{R19, R20}.


Consider a scenario where we have a dataset with dimensionality $N$, and we aim to perform PCA to reduce the dimensionality to $n$, where $n < N$. Let $\theta = [\theta_{1}, \theta_{2}, \dots, \theta_{N}]$ represent the original data vector. We seek to find $n$ principal components that capture the most variance in the dataset. These principal components are represented by the eigenvectors of the covariance matrix of the dataset. Let $P_{i}^{T}$ denote the transpose of the eigenvector matrix, where $i = 1, 2, \dots, n$. The eigenvector matrix contains the eigenvectors corresponding to the $n$ largest eigenvalues. To transform the original data into the principal component space, we compute the dot product of $\theta$ with each eigenvector in $P_{i}^{T}$. This operation yields the transformed vector $\phi = [\phi_{1}, \phi_{2}, \dots, \phi_{n}]$. By selecting $n$ principal components, we achieve dimensionality reduction while retaining the most important information in the dataset. The transformed vector $\phi$ now represents the data in the reduced-dimensional principal component space
\begin{equation}
    \phi_{i} = P_{i}^{T} \theta_{i} ,
    \label{eq:pca_general_form}
\end{equation}
we have the general form of the projection onto a principal component $i$, represented by $\phi_{i}$. This projection is achieved through the dot product between $P_{i}^{T}$ and $\theta_{i}$, the original data vector.
\begin{equation}
   (\phi_{1})\geq (\phi_{2}) \geq \dots \geq (\phi_{i}) > 0
    \label{eq:pca_variance}
\end{equation}

The variance of the principal components follows the relationship described by Eq.(\ref{eq:pca_variance}). As the index increases, the variance of the principal components generally decreases. Consequently, higher index values correspond to a reduced amount of variance information contained in the data.

\subsection{t-SNE}

The t-SNE is a method used for data visualization and dimensionality reduction, similar to PCA. However, the key distinction lies in its capability to capture nonlinear relationships within the data and preserve the high-dimensional structure when mapping it to a lower-dimensional space \cite{R21, R22}. Unlike PCA, which focuses on capturing variance, t-SNE emphasizes on maintaining the local structure and relationships between data points. This makes t-SNE a valuable tool for visualizing complex data patterns and uncovering hidden clusters or groupings that may not be readily apparent in the original high-dimensional space.


The t-SNE algorithm utilizes a Gaussian kernel to establish pairwise similarities by measuring the distances between points in the original dataset. This generates probability distributions over pairs of points, where the likelihood of being similar is determined by their pairwise similarity. Subsequently, the algorithm maps these selected points to a lower-dimensional space, aiming to create a similar probability distribution. The objective of t-SNE is to minimize the discrepancy between these two distributions, ultimately yielding a lower-dimensional representation that preserves the underlying structure of the original data \cite{R21, R22}. By capturing nonlinear relationships and preserving the local structure, t-SNE enables the visualization of intricate patterns and clusters that may be hidden in the higher-dimensional space. The pairwise probability is defined as
\begin{equation}
    p_{ij} = \frac{p_{j|i} + p_{i|j}}{2n}
    \label{eq:tsne_pij},
\end{equation}
where $p_{j|i}$ and $p_{i|j}$ represent the likelihood of point $j$ given point $i$ and vice versa, respectively. The total number of points in the dataset is denoted as $n$. Notably, in t-SNE, $p_{ii}$ and $p_{jj}$ are set to zero, and $p_{ij} = p_{ji}$. The calculation of these probabilities is determined by Eq.(\ref{eq:tsne_pij}), which generates the pairwise similarities in the original space.
Then:
\begin{equation}
    q_{ij} = \frac{\left (1+ \left \| y_{i} - y_{j} \right \|^{2}  \right )^{-1}}{\sum_{k \neq l}^{}\left (1+ \left \| y_{k} - y_{l} \right \|^{2}  \right )^{-1}} ,
    \label{eq:tsne_qij}
\end{equation}
is applied to create map candidates, denoted as $y = y_{1}, y_{2}, \dots, y_{n}$, in a lower-dimensional space. Initially, these candidates are randomly initialized, often by sampling from a Gaussian distribution with a small variance centered at the origin. The objective of t-SNE is to determine the optimal mapping relationships between the high-dimensional and low-dimensional spaces. To achieve this, t-SNE minimizes the Kullback-Leibler divergence, a measure of dissimilarity between probability distributions. 
\begin{equation}
    KL(P||Q) = \sum_{ij}^{} p_{ij}\textup{log}\left ( \frac{p_{ij}}{q_{ij}} \right ).
    \label{eq:kb_divergence}
\end{equation}

\section{Problems to analyze}

As mentioned earlier, the central objective of this paper is to explore the utilization of the QAOA for solving max-cut problems. In particular, our focus is on examining the effects of employing entangled and non-entangled mixing operators at different depths \cite{R10}. The max-cut problems analyzed in our study are characterized by specific configurations of phase and mixing operators, which can be categorized as cyclic or complete configurations. These configurations play a crucial role in shaping the behavior and performance of the QAOA in tackling max-cut problems. In the max-cut problem with the cyclic configuration, the phase operator can be represented as
\begin{equation}
    U(H_{cyc}, \gamma) = e^{-i\gamma H_{cyc}} = \prod_{\langle j, k \rangle}^{} e^{-i\gamma Z_{j}Z_{k}}
    \label{eq:po_mc_cyclic}
\end{equation}
In this Eq.(\ref{eq:po_mc_cyclic}), the notation $\langle j, k \rangle$ signifies the connection between neighboring nodes. For instance, in the context of a 4-node problem, Eq.(\ref{eq:po_mc_cyclic}) captures the connections between node 1 and node 2, node 2 and node 3, node 3 and node 4, and node 4 and node 1. It is important to note that the final connection completes the cycle, ensuring that all nodes are interconnected in a cyclic manner. This representation of the phase operator is fundamental in formulating the QAOA for solving the max-cut problem with the cyclic configuration. In the max-cut problem with the complete configuration, the phase operator is represented by Eq.(\ref{eq:po_mc_complete}).
\begin{equation}
    U(H_{com}, \gamma) = e^{-i\gamma H_{com}} = \prod_{ \left \{ j, k \mid j \neq k \right \} }^{} e^{-i\gamma Z_{j}Z_{k}}
    \label{eq:po_mc_complete}
\end{equation}

In this configuration, every pair of nodes in the graph is connected, excluding self-connections ($\left\{ j, k \mid j \neq k \right\}$). Notably, connections between nodes are not repeated, meaning that a connection from node $j$ to node $k$ is considered the same as a connection from node $k$ to node $j$. This absence of directionality in the max-cut problem leads to a symmetric representation of the phase operator, reflecting the undirected nature of the connections in the complete configuration. Eq.(\ref{eq:po_mc_complete}) provides the fundamental formulation for the phase operator in the QAOA for solving the max-cut problem with the complete configuration. The mixing operator without entanglement for both max-cut configurations is expressed by  
\begin{equation}
    U(H_{B}, \beta_{1}, \beta_{2}) = e^{i \beta_{1} \beta_{2} H_{B}} = \prod_{ j }^{} e^{i\beta_{1} X_{j}} e^{i\beta_{2} Y_{j}} ,
    \label{eq:mo_mc_nonentangled}
\end{equation}
In this equation (\ref{eq:mo_mc_nonentangled}), the mixing operator incorporates $RX$ and $RY$ rotations, setting it apart from the conventional mixing operators typically employed.


The entangled mixing operator incorporates an additional term compared to the non-entangled case. This term is generated by applying $CNOT$ gates between each qubit (node) in the system, akin to the complete configuration. The equation representing the entangled mixing operator is given by
\begin{equation}
    e^{i \frac{\pi}{4} \left ( I - Z \right ) \otimes \left ( I - X \right )} = \begin{pmatrix}
            I & 0 \\ 
            0 & X
            \end{pmatrix} = 
    \begin{pmatrix}
            1 & 0 & 0 & 0 \\ 
            0 & 1 & 0 & 0 \\ 
            0 & 0 & 0 & 1 \\ 
            0 & 0 & 1 & 0
            \end{pmatrix} ,
    \label{eq:cnot_ent_mo}
\end{equation}
to represent the entangled mixing operator in our notation. We utilize the expression $e^{i I_{j}X_{k}}$ to indicate that qubit $j$ (node) acts as the control while qubit $k$ (node) serves as the target. By incorporating the term $e^{i I_{j}X_{k}}$ into the previous mixing operator without entanglement, we obtain the following expression:
\begin{equation*}
    U(H_{B}, \beta_{1}, \beta_{2}) = e^{i \beta_{1} \beta_{2} H_{B}}
\end{equation*}
\begin{equation}
    = \prod_{ j }^{} e^{i\beta_{1} X_{j}} \prod_{ \left \{ j, k \mid j \neq k \right \} }^{} e^{i I_{j}X_{k}} \prod_{ j }^{}e^{i\beta_{2} Y_{j}} ,
    \label{eq:mo_mc_entangled}
\end{equation}
which represents the mixing operator with an entanglement stage inserted between the $RX$ and $RY$ rotations. The entanglement stage facilitates interactions between each pair of qubits in the system, guaranteeing that there are no duplicated interactions and excluding self-interactions. This entanglement stage enhances the connectivity and entanglement among the qubits, contributing to the overall behavior and performance of the mixing operator in the quantum system.

\begin{figure}[ht]
    \begin{center}
        \begin{tikzpicture}
            \node[scale=0.65] {
                \begin{quantikz}
                    \ket{\psi_{1}}  & \gate{RX} & \qw & \gate{RY} & \qw \\
                    \ket{\psi_{2}} & \gate{RX} & \qw & \gate{RY} & \qw \\
                    \ket{\psi_{3}} & \gate{RX} & \qw & \gate{RY} & \qw \\
                \end{quantikz}
            };
        \end{tikzpicture}
    \end{center}
    \caption{Individual rotations in mixing operators.}
    \label{dia:mixing_operator_indivual_rotations}
\end{figure}
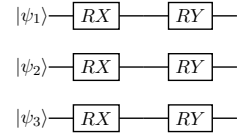

\begin{figure}[ht]
    \begin{center}
        \begin{tikzpicture}
            \node[scale=0.65] {
                \begin{quantikz}
                    \ket{\psi_{1}}  & \gate{RX} & \ctrl{1} & \ctrl{2} & \qw & \gate{RY} & \qw \\
                    \ket{\psi_{2}} & \gate{RX} & \targ{} & \qw & \ctrl{1} & \gate{RY} & \qw \\
                    \ket{\psi_{3}} & \gate{RX} & \qw & \targ{} & \targ{} & \gate{RY} & \qw \\
                \end{quantikz}
            };
        \end{tikzpicture}
    \end{center}
    \caption{Entangled rotations in mixing operators.}
    \label{dia:mixing_operator_entangled_rotations}
\end{figure}
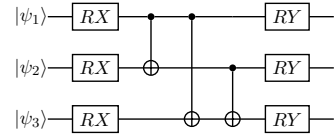

The quantum circuit representation of the two types of tested mixing operators can be seen in Figures \ref{dia:mixing_operator_indivual_rotations} and \ref{dia:mixing_operator_entangled_rotations}. {Figure \ref{dia:mixing_operator_indivual_rotations}} corresponds to the non-entangled mixing operator, while {Figure \ref{dia:mixing_operator_entangled_rotations}} corresponds to the entangled mixing operator.

In our analysis, both PCA and t-SNE methods were applied to each QAOA model, namely the $1L$, $2L$, and $3L$ models. The $1L$ model consists of one phase operator with an associated parameter $\gamma$ and one mixing operator with two associated parameters $\beta_1$ and $\beta_2$. The $2L$ model includes two phase operators with parameters $\gamma_1$ and $\gamma_2$, along with two mixing operators (connected between each phase operator) with parameters $\beta_{1-1}$, $\beta_{1-2}$, $\beta_{2-1}$, and $\beta_{2-2}$. The $3L$ model, although tested for some problems only, comprises three-phase operators with parameters $\gamma_1$, $\gamma_2$, and $\gamma_3$, and three mixing operators with a total of six associated parameters.

For the application of PCA and t-SNE methods, we utilized the best result from each experiment and performed a total of 100 experiments for each problem. Specifically, the parameters (angles) from the best results were used to create a dataset. For example, in the case of the $1L$ QAOA model (with and without entanglement), we created a dataset with three columns representing the parameters $\gamma$, $\beta_1$, and $\beta_2$, and 100 rows corresponding to the experiments, resulting in a dimension of $3 \times 100$. This process was repeated for all models and problems, with the number of dimensions varying accordingly ($3 \times 100$ for $1L$, $6 \times 100$ for $2L$, and $9 \times 100$ for $3L$). The entanglement between similar models was also considered in the analysis.


The dataset used in our analysis consists of the optimal parameters obtained for the phase ($RZ$ gates) and mixing operators ($RX$ and $RY$ gates) using the QAOA with Stochastic Hill Climbing with Random Restarts (SHC-RR) as the optimization method. SHC-RR is known for its unbiased exploration of the search space, making it a suitable strategy for generating diverse data.

We applied the QAOA with SHC-RR to solve max-cut problems with cyclic and complete configurations, considering different numbers of nodes: 4 nodes (4n), 10 nodes (10n), and 15 nodes (15n). Each problem was simulated 100 times to account for variability. For each problem instance, we evaluated three different QAOA depths: $1L$, $2L$, and $3L$. Additionally, each QAOA model was assessed with and without an entanglement stage in the mixing operators.

By combining the QAOA optimization process with the SHC-RR method, we generated a comprehensive dataset of optimal parameters for further analysis and visualization using PCA and t-SNE techniques.

\begin{figure}[ht]
\centering
\includegraphics[width=8cm, height=6cm]{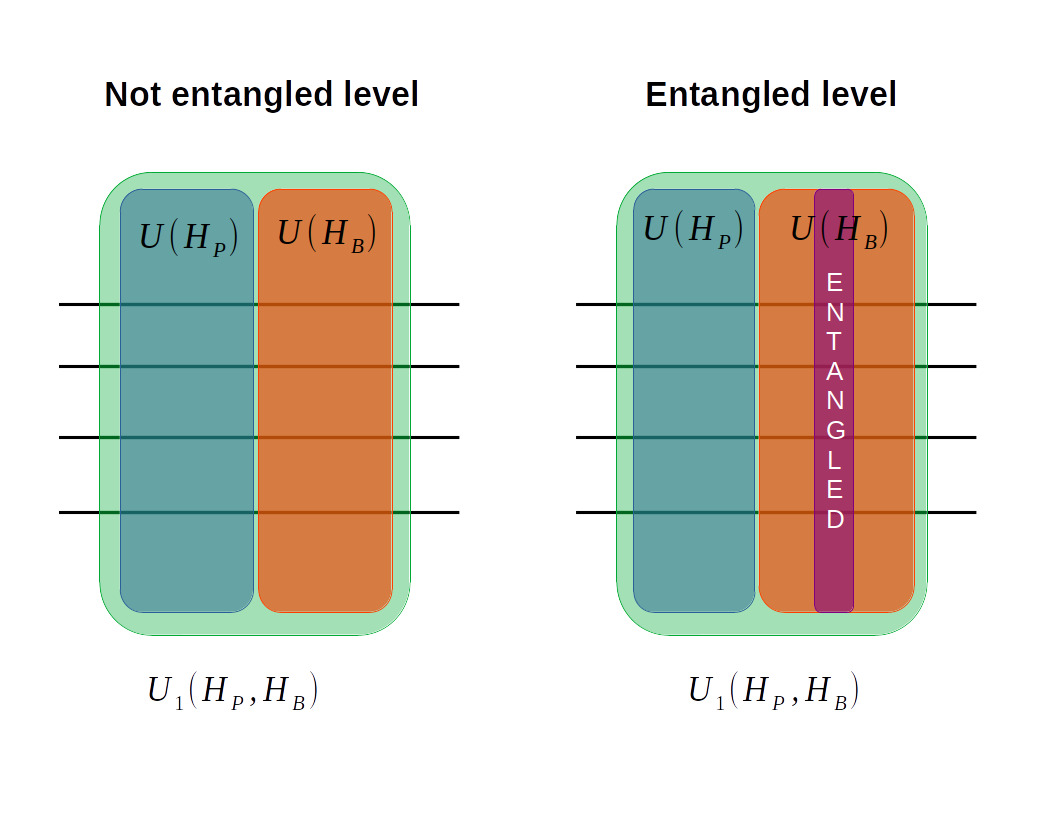}
\caption{Non-entangled and entangled mixing operators.}
\label{fig:ent_and_notent}
\end{figure}

The dataset of solved max-cut problems used in this study was generated specifically for our forthcoming research, which includes a comprehensive analysis of the QAOA. For detailed results and information on all the experiments performed, including the optimized parameters, we refer the reader to reference \cite{R23}. It is important to note that for the 10n and 15n problems, we performed experiments using QAOA depths of $1L$, $2L$, and $3L$ to explore the algorithm's performance in more complex scenarios. However, for the 4n problems, which are relatively simpler in nature, we focused on depths of $1L$ and $2L$.

\begin{figure}[ht]
\centering
\includegraphics[width=8cm, height=6cm]{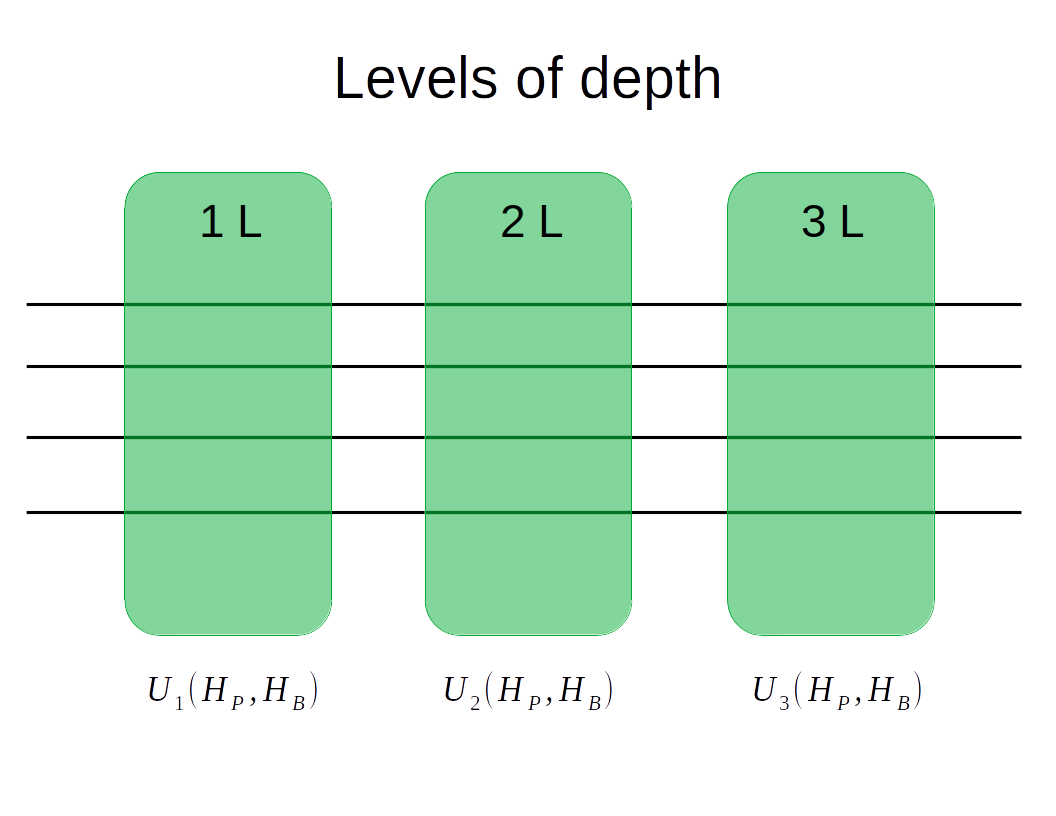}
\caption{Levels of depth, with one pair of phase and mixing operator $1L$, two pairs $2L$ and three pairs $3L$.}
\label{fig:levels_depth}
\end{figure}

\section{Experiments and results}

The PCA method is employed to identify correlations and patterns within a dataset by analyzing the variance among data points. By capturing the most significant sources of variance, PCA allows for the reduction of the original attributes into a new basis with fewer dimensions. These new dimensions, known as principal components, aim to retain as much information as possible from the original data.

In our study, we applied the PCA method to analyze the entangled and non-entangled QAOA models individually, as well as for each pair of models with compatible dimensions that differed only in the presence of the entanglement stage in the mixing operator. By performing PCA on these models, we aimed to explore and compare the variance and information contained in their respective principal components. This analysis provides insights into the impact of the entanglement stage on the overall behavior and performance of the QAOA models.


The t-SNE method, similar to PCA, aims to reduce the dimensionality of the data and provide insights into the relationships between data points. However, t-SNE utilizes a different approach to accomplish this. It models the data as a pairwise probability distribution, where each distribution represents the likelihood of two data points being related or similar. By iteratively minimizing the Kullback-Leibler divergence, which is a measure of the difference between two probability distributions, t-SNE finds a lower-dimensional representation of the data while preserving its intrinsic structure.

During the t-SNE process, the algorithm focuses on preserving the local relationships between data points rather than the global structure. This makes t-SNE particularly effective in capturing nonlinear relationships and clustering patterns in the data. By examining the resulting embedding in the lower-dimensional space, we can gain insights into the similarities and dissimilarities between data points.

In our experiments, we applied t-SNE after optimizing the QAOA parameters to obtain a low-dimensional representation of the data. We also calculated the Kullback-Leibler divergence, which measures the amount of information loss during the embedding process. A lower divergence value indicates a better representation of the high-dimensional data, while a higher value suggests a significant loss of information. By analyzing the divergence values, we can evaluate the quality of the t-SNE embedding and its ability to capture the important features of the QAOA models.


\subsection{PCA applied to QAOA problems}

In our analysis, we applied PCA to each dataset in two steps: individual PCA analysis and paired PCA analysis. In the individual PCA analysis, we performed PCA on each dataset separately, considering the first three principal components. This allowed us to explore the variance and structure of the data within each model independently. Additionally, we calculated the explained variance associated with each principal component, which provided insights into the amount of information captured by each component. In the paired PCA analysis, we compared the individual PCA projections of models with the same number of rotation parameters (dimensions). Specifically, we selected the first three principal components of both models and examined how the combined PCA projection differed from the original individual projections. This comparison allowed us to assess the variations between the projections and gain a deeper understanding of the relationships between models.

It is worth noting that we applied the individual and paired PCA approaches to all three levels of depth (1L, 2L, and 3L) in order to ensure a fair basis for comparison between models. For the 1L depth level, which corresponds to the 3-parameter model (3p), PCA is not primarily used for dimensionality reduction since the number of parameters matches the number of PCA components we are considering. However, employing PCA, in this case, allows us to identify correlations between the parameters, providing insights into the relative importance of certain gates within the QAOA. For the 2L and 3L depth levels, corresponding to the 6-parameter (6p) and 9-parameter (9p) models, respectively, the first three PCA components serve both as indicators of parameter correlations within the QAOA and for dimensionality reduction purposes.

\begin{table}[ht]
\footnotesize
\centering
\begin{tabular}{|c|c|c|c|}
\hline
Parameters & PCA 1      & PCA 2      & PCA 3  \\ \hline
3 p         & 0.44317179 & 0.29359184 & 0.26323637 \\ \hline
3 p ent     & 0.4189022  & 0.30327991 & 0.2778179 \\ \hline
6 p         & 0.23918661 & 0.20909359 & 0.16835319 \\ \hline
6 p ent     & 0.2949565  & 0.22615379 & 0.1985824 \\ \hline
\end{tabular}
\caption{Individual PCA projections explained variance (4n cyclic) for the first 3 PCA components.}
\label{tab:exp_var_indPCA_4nCYC}
\end{table}

In {Table \ref{tab:exp_var_indPCA_4nCYC}}, we present the explained variances for the individual PCA projections of the 4n cyclic configuration max-cut problem. For the first 1L model (3 parameters), we observed a decrease in correlation in the first PCA component for the entangled model compared to the non-entangled model. This decrease suggests a reduced significance of this component in terms of representation importance. However, the second and third PCA components showed an increase in explained variance for the entangled model, indicating an increased contribution to the overall variance.
Conversely, for the 6-parameter models (2L and 3L), the phenomenon is reversed. The entangled models demonstrated higher explained variance values for the first PCA component, indicating its increased significance in capturing variance. When comparing the first three PCA components, all variance values were greater in the entangled models, suggesting that the entangled mixing operator contributed to a higher total amount of variance contained in the components.

These observations highlight the impact of entanglement on the correlation structure and the distribution of explained variance within the PCA components for different QAOA models.

\begin{figure}[ht]
\centering
\includegraphics[width=8cm, height=6cm]{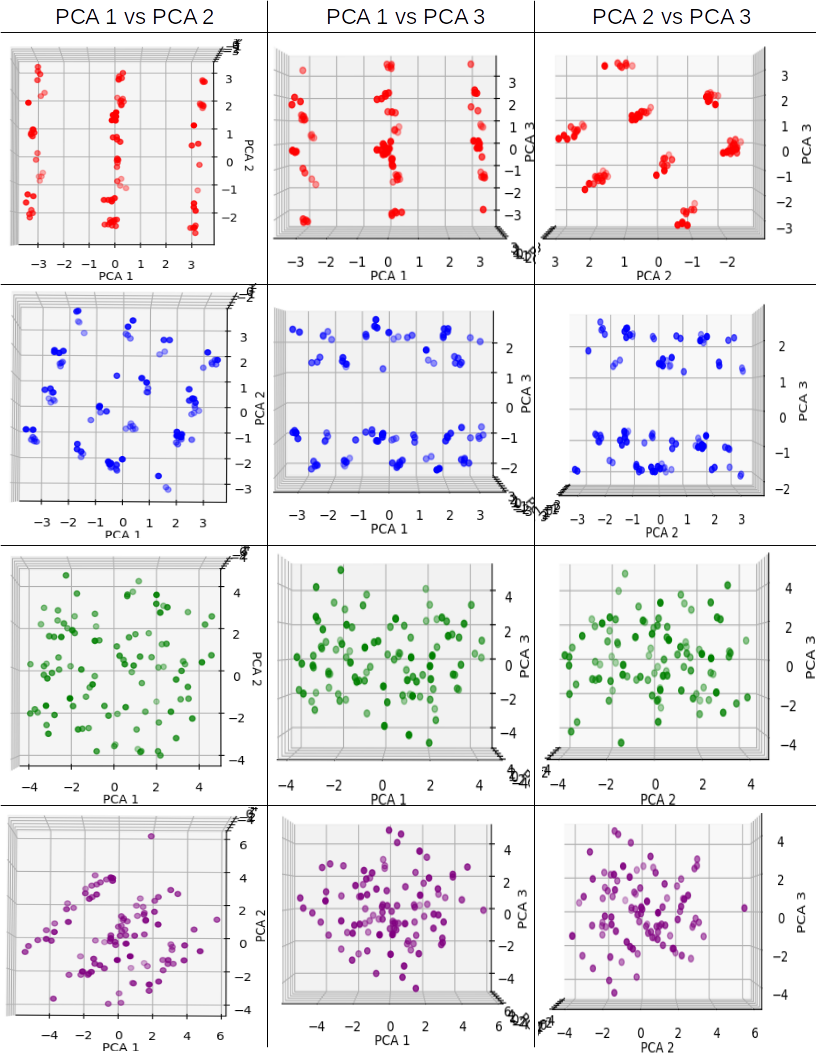}
\caption{PCA individual graphs for 4n cyclic configuration max-cut problem solved using QAOA, first 3 components. Red corresponds to the $3p$ parameter $1L$ non-entangled, blue $3p$ parameter $1L$ entangled, green $6p$ parameter $2L$ non-entangled and purple $6p$ parameter $2L$ entangled model.}
\label{fig:comp_ind_pca_4nCYC}
\end{figure}

In {Figure \ref{fig:comp_ind_pca_4nCYC}}, we present the individual PCA graphs for the 4-node cyclic max-cut problem. For the red model (non-entangled $3p$), the first two graphs (PCA 1 vs PCA 2 and PCA 1 vs PCA 3) exhibit distinct line patterns. These patterns suggest correlations and structure in the data. In contrast, the blue model (entangled $3p$) shows similar line behavior in the PCA 1 vs PCA 2 graph, but with an additional line compared to the non-entangled model. In the PCA 1 vs PCA 3 and PCA 2 vs PCA 3 graphs, a separation pattern with two distinct groups can be observed. In the $6p$ parameter ($2L$ depth) models, the green model (non-entangled) does not exhibit any recognizable patterns or clusters in the graphs. On the other hand, the purple model (entangled) shows three distinct cluster lines in the PCA 1 vs PCA 2 graph. However, no recognizable patterns are observed in the remaining graphs.

These observations provide insights into the correlation structures and clustering tendencies present in the individual PCA projections for the different QAOA models. 
\begin{table}[ht]
\centering
\footnotesize
\centering
\begin{tabular}{|c|c|c|c|}
\hline
Parameters & PCA 1      & PCA 2      & PCA 3 \\ \hline
3 p         & 0.42701248 & 0.29829486 & 0.27469266 \\ \hline
6 p         & 0.22597391 & 0.19226576 & 0.18187536 \\ \hline
\end{tabular}
\caption{Pair PCA projections explained variance for the first 3 PCA components for the 4n cyclic max-cut problem.}
\label{tab:exp_var_pairPCA_4nCYC}
\end{table}

In the case of using a pair PCA model in the 4n cyclic max-cut problem for the $3p$ and $6p$, the results for the PCA components are shown in {Table \ref{tab:exp_var_pairPCA_4nCYC}}. In the case of the $3p$ pair model, the explained variance values of the PCA components demonstrate an intermediate trend between the entangled and non-entangled models from the separate PCA analysis. This suggests that the pair PCA model captures a combination of characteristics from both individual models, resulting in a representation that lies between the entangled and non-entangled configurations. For the $6p$ pair model, the variance of the PCA components appears to be closer to the values of the $6p$ non-entangled model from the separate PCA analysis. This indicates that the pair PCA model aligns more closely with the non-entangled configuration in terms of the distribution of variance across the PCA components. However, it is worth noting that the sum of the variances for the first three components in the $6p$ pair model accounts for only 60$\%$ of the total information variance. This suggests that the pair PCA model for the $6p$ parameter configuration retains a relatively low amount of original information in the new representation. As a result, it becomes more challenging to identify specific trends or patterns in the pair PCA projection for this type of model, given the reduced amount of original information maintained.


\begin{figure}[ht]
\centering
\includegraphics[width=8cm, height=6cm]{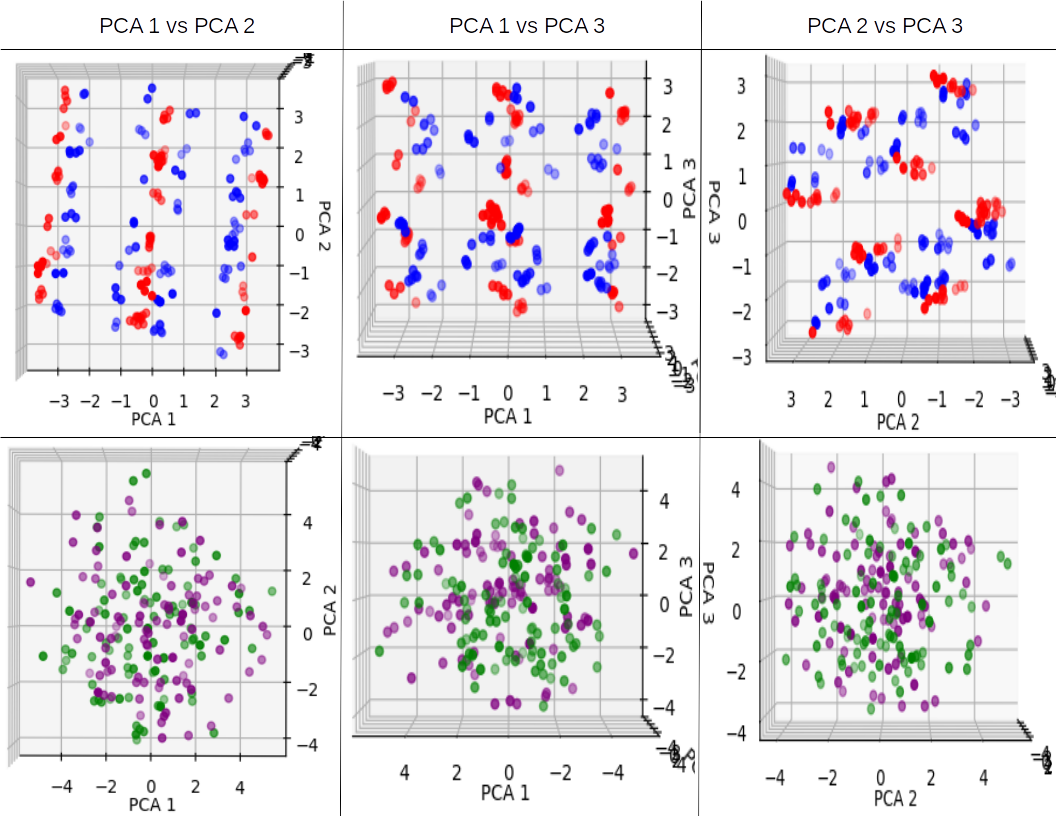}
\caption{PCA pair graphs for 4n cyclic configuration max-cut problem solved using QAOA, first 3 components. Red corresponds to the $3p$ parameter $1L$ non-entangled, blue $3p$ parameter $1L$ entangled, green $6p$ parameter $2L$ non-entangled and purple $6p$ parameter $2L$ entangled model.}
\label{fig:comp_pair_pca_4n_CYC}
\end{figure}

In {Figure \ref{fig:comp_pair_pca_4n_CYC}}, for the $3p$ and $6p$ configurations, we observe different behaviors. For the $3p$ models (red non-entangled and blue entangled), the patterns observed in the individual PCA graphs are preserved. However, when both entangled and non-entangled models are combined in the pair PCA analysis, we can see that the data points of the models are projected into different areas while still following the same patterns observed in the previous graphs. This suggests that the entanglement stage in the mixing operator introduces a noticeable shift in the distribution of the projected points, resulting in different regions for the entangled and non-entangled models.

In contrast, for the $6p$ models, the previous patterns observed in the individual PCA graphs do not hold. The distribution of the projected points appears to be random in the majority of the pair PCA graphs. Only in the PCA 1 vs PCA 3 graph, we can observe some pattern with small centered line clusters for the entangled model (purple), while the non-entangled model (green) exhibits a more scattered distribution compared to the purple data. This suggests that for the $6p$ parameter configuration, the entanglement stage in the mixing operator does not introduce a clear and consistent pattern in the pair PCA projections, leading to a more random distribution of points.

\subsection{t-SNE applied to QAOA problems}


In our analysis, we employed t-SNE as an additional method to gain insights into the relationships within the QAOA data from the solved max-cut problems. By utilizing multiple tools, we aimed to extract valuable information about the entanglement stage and investigate its impact on the overall data relationships.

To perform the t-SNE analysis, we experimented with different perplexity values. Perplexity is a crucial parameter that determines the number of nearest neighbors considered during the construction of the lower-dimensional representation. We tested three perplexity values: $3$, $30$ (the default value in the Sklearn package), and $99$ (an additional value of $199$ specifically for pair models). The purpose of exploring different perplexity levels was to identify potential variations in data behavior under different settings.

For the initialization of the t-SNE embedding, we employed PCA. This choice of initialization using PCA helps to provide a more globally stable solution compared to the random initialization. By using a consistent initialization method, we aimed to ensure a more accurate and precise comparison between the models.

By employing t-SNE with varying perplexity values and initializing with PCA, we aimed to uncover hidden patterns and relationships in the QAOA data, with a particular focus on the entanglement stage and its influence on the data structure.


Also, as in the case of PCA, we created individual and paired t-SNE graphs for each problem and for each model using $3p$ parameters, $6p$ parameters, and $9p$ parameters (for the 10n and 15n problems). These t-SNE graphs allowed us to explore the data distributions and identify patterns and clusters within each specific model and problem. The individual t-SNE graphs provided insights into the intrinsic structures of the data, while the paired t-SNE graphs facilitated a direct comparison between the entangled and non-entangled models, revealing any differences or similarities in their data relationships.


\begin{table}[ht]
\centering
\footnotesize
\begin{tabular}{|c|ccc|}
\hline
           & \multicolumn{3}{c|}{KL - Divergence}                                           \\ \hline
Parameters & \multicolumn{1}{c|}{KL-D}       & \multicolumn{1}{c|}{KL-D}       & KL-D       \\ \hline
3 p        & \multicolumn{1}{c|}{0.12658161} & \multicolumn{1}{c|}{0.18968964} & 0.00003131 \\ \hline
3 p ent    & \multicolumn{1}{c|}{0.08551478} & \multicolumn{1}{c|}{0.20780504} & 0.00004092 \\ \hline
6 p        & \multicolumn{1}{c|}{0.55863631} & \multicolumn{1}{c|}{0.4951154}  & 0.00003326 \\ \hline
6 p ent    & \multicolumn{1}{c|}{0.35603735} & \multicolumn{1}{c|}{0.39002356} & 0.00004086 \\ \hline
\end{tabular}
\caption{Individual KL-Divergence for 4n cyclic max-cut problem with different numbers of perplexity, considering the $3p$ non-entangled, $3p$ entangled, $6p$ non-entangled and $6p$ entangled models.}
\label{tab:ind_kl_divergence_t-SNE}
\end{table}


For the first problem, at a perplexity of $3$, the non-entangled models had higher (worse) KL-D values after the mapping compared to the entangled models. At a perplexity of $30$, the $3p$ non-entangled model had a better (lower) KL-D value compared to the entangled model, and for the $6p$ models, the entangled model still had a better KL-D value compared to the non-entangled one. Finally, at $99$ perplexity, all models showed good KL-D values, indicating a better mapping for all the perplexity values tested. In this case, the non-entangled and entangled models had a slight difference in performance, with the non-entangled models performing slightly better at this perplexity, although the difference can be considered negligible

\begin{figure}[ht]
\centering
\includegraphics[width=8cm, height=6cm]{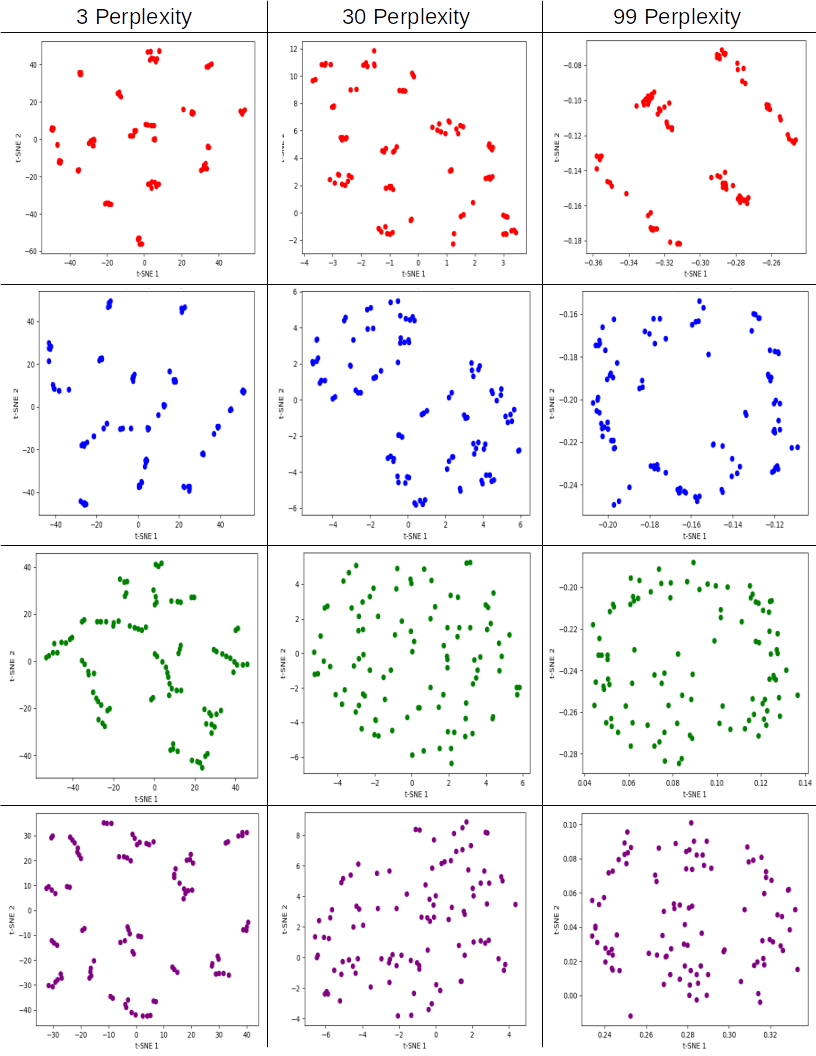}
\caption{t-SNE individual graphs for 4n cyclic configuration max-cut problem solved using QAOA, with different perplexity values $3$, $30$ and $99$. Red corresponds to the $3p$ parameter $1L$ non-entangled, blue $3p$ parameter $1L$ entangled, green $6p$ parameter $2L$ non-entangled and purple $6p$ parameter $2L$ entangled model.}
\label{fig:comp_ind_t-SNE_4nCYC}
\end{figure}

The individual t-SNE graphs for the 4n cyclic max-cut problem are shown in {Figure \ref{fig:comp_ind_t-SNE_4nCYC}}. For the $3p$ non-entangled model (red), the most significant pattern can be observed in the $99$ perplexity level, which has a linear pattern similar to the one obtained in the PCA graph for the model and problem. For the $3p$ entangled model (blue), the $30$ perplexity level shows a two-cluster pattern, and the $99$ perplexity level shows a circular pattern with no data points in the center of the plane. For the $6p$ non-entangled model (green), the $30$ perplexity level has a similar distribution to the one obtained in the PCA individual graph for the same model, with a random distribution pattern. Finally, for the $6p$ entangled model (purple), the most significant pattern can be observed in the $99$ perplexity level, which has an external circle with a middle line pattern.


\begin{table}[ht]
\centering
\footnotesize
\begin{tabular}{|c|cc|}
\hline
          & \multicolumn{2}{c|}{KL-Divergence}           \\ \hline
Parameter & \multicolumn{1}{c|}{3 per}      & 30 per     \\ \hline
3 p       & \multicolumn{1}{c|}{0.11827804} & 0.24608216 \\ \hline
6 p       & \multicolumn{1}{c|}{0.58829921} & 0.73377264 \\ \hline
          & \multicolumn{1}{c|}{99 per}     & 199 per    \\ \hline
3 p       & \multicolumn{1}{c|}{0.17905515} & 0.00003183 \\ \hline
6 p       & \multicolumn{1}{c|}{0.33970055} & 0.00004709 \\ \hline
\end{tabular}
\caption{Pair KL-Divergence for 4n cyclic max-cut problem with different numbers of perplexity, considering the $3p$ parameters (non-entangled and entangled) and $6p$ parameters (non-entangled and entangled) models.}
\label{tab:pair_kl_divergence_t-SNE}
\end{table}

The KL-D values for the pair models in the 4n cyclic max-cut problem are presented in {Table \ref{tab:pair_kl_divergence_t-SNE}}. Interestingly, all the best KL-D values were obtained by the $3p$ models. Consistent with the individual t-SNE analysis, the best KL-D values were obtained with the highest perplexity value.

\begin{figure}[ht]
\centering
\includegraphics[width=8cm, height=6cm]{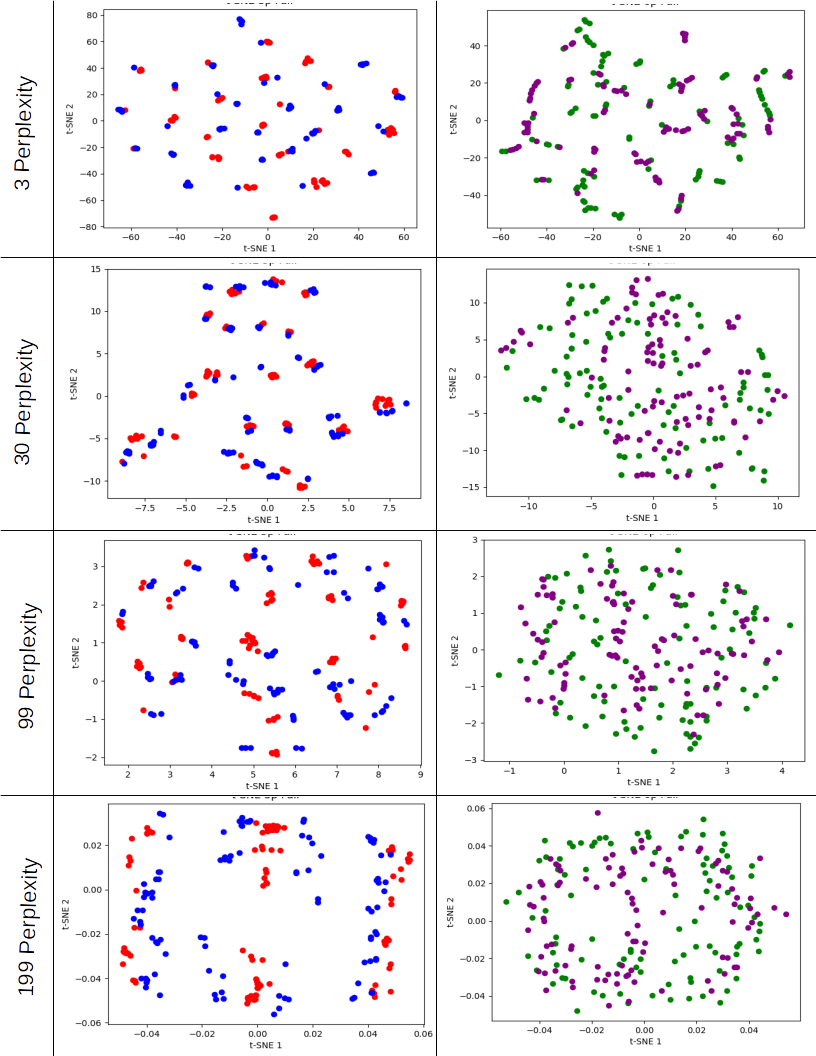}
\caption{t-SNE pair graphs for 4n cyclic configuration max-cut problem solved using QAOA, with different perplexity values $3$, $30$, $99$ and $199$. Red corresponds to the $3p$ parameter $1L$ non-entangled, blue $3p$ parameter $1L$ entangled, green $6p$ parameter $2L$ non-entangled and purple $6p$ parameter $2L$ entangled model.}
\label{fig:comp_pair_t-SNE_4nCYC}
\end{figure}

In the pair t-SNE model graphs ({Figure \ref{fig:comp_pair_t-SNE_4nCYC}}) for the 4n cyclic max-cut problem, we start by focusing on the $3p$ models non-entangled and entangled (red and blue, respectively) at $199$ perplexity. The line patterns of the red model are maintained, but the blue model shows a completely different distribution, where it has a similar pattern as the red model. The entangled model data contains the red points at the center, but at the extremes, the red model seems to contain the blue data. For the $6p$ models non-entangled and entangled (green and purple, respectively), interesting results are seen at the $99$ and $199$ perplexity values. The purple model tends to be grouped in certain areas of the plane in the $99$ perplexity, while the green model has a random distribution in the plane with no particular pattern. For the $199$ perplexity, the patterns seen in the individual graphs are maintained, with an elliptical behavior, and in particular, the purple model shows a line pattern at the center.

\subsection{Results analysis}


In the analysis performed using the PCA method, it is noteworthy that the $3p$ models (corresponding to the $1L$ depth) for both cyclic and complete max-cut problems consistently demonstrate the highest explained variance in the first 3 components. This can be attributed to the fact that these models have the same number of parameters (dimensions) as the number of PCA components, resulting in no dimensionality reduction and no loss of information. This characteristic distinguishes these models from others and highlights the correlations between specific gates within the QAOA. However, further investigations are necessary to identify the precise interactions between gates that contribute significantly to the observed variances in specific PCA components. The graphical representations of the $3p$ models consistently exhibit linear patterns, suggesting a certain consistency in the relationships between the parameter values and the problem type they originate from. These findings indicate the existence of underlying structure and potential optimization strategies within the QAOA framework. Nevertheless, it is important to perform additional studies to unravel the specific gate interactions and their implications for problem-solving capabilities. By gaining a deeper understanding of the influential gate correlations, it may be possible to enhance the performance and efficiency of the QAOA in tackling max-cut and other combinatorial optimization problems.

In the analysis of the $6p$ models (corresponding to $2L$ depth) using PCA, intriguing behaviors emerge due to the increased complexity of these models. The dimensionality reduction provided by the first 3 PCA components proves to be a valuable strategy in capturing essential information. By examining the explained variances in the individual PCA graphs for the cyclic configuration problems of 4n, 10n, and 15n (refer to Tables \ref{tab:exp_var_indPCA_4nCYC}, \ref{tab:exp_var_indPCA_10nCYC}, and \ref{tab:exp_var_indPCA_15nCYC}), we consistently observe that the entangled models exhibit higher values in the PCA components. This consistent trend between the entangled and non-entangled models indicates a discernible difference, suggesting that the presence of the entanglement stage in the mixing operator enriches the QAOA parameters with a greater amount of information (variance) that can be detected through PCA analysis.

From a graphical standpoint, when focusing on the PCA 1 vs PCA 2 plane, which contains the most relevant and informative data as shown in Figures \ref{fig:comp_ind_pca_4nCYC}, \ref{fig:comp_ind_pca_10nCYC}, and \ref{fig:comp_ind_pca_15nCYC}, distinct differences emerge between the non-entangled (green) and entangled (purple) models. In the case of the non-entangled models, the mapped data exhibits a seemingly random distribution, likely attributed to the individual rotations within the model. On the other hand, the entangled models display clear clustering behaviors, leading to visually discernible disparities in the graphs. It is worth emphasizing that the PCA method remains oblivious to the fact that the processed data originates from an entangled circuit; however, it effectively captures and represents the inherent distinctions in the data distribution.

Furthermore, in the case of the $6p$ models for the complete configuration problems, specifically the 4n and 15n problems (refer to Tables \ref{tab:exp_var_indPCA_4nCOM}, \ref{tab:exp_var_indPCA_10nCOM}, and \ref{tab:exp_var_indPCA_15nCOM}), we observe similar behaviors to those observed in the cyclic problems. The entangled models consistently exhibit higher values in the PCA components compared to the non-entangled models. Additionally, for the 10n complete problem, the entangled model demonstrates a slightly higher total explained variance compared to the non-entangled model. Examining the graphical representations (refer to Figures \ref{fig:comp_ind_pca_4nCOM}, \ref{fig:comp_ind_pca_10nCOM}, and \ref{fig:comp_ind_pca_15nCOM}), most of the entangled models exhibit clustering behaviors, while the non-entangled models do not show a clear pattern or distribution. These observations further highlight the distinguishing characteristics between entangled and non-entangled models in terms of their PCA representations.

In the case of the $9p$ models (corresponding to $3L$ depth) for both cyclic and complete configurations, we once again observe higher PCA values for the entangled models, regardless of the problem type. However, it is important to note that the total amount of variance in the $9p$ models is relatively low. Consequently, when examining the graphical representations (refer to Figures \ref{fig:comp_ind_pca_10nCYC_2}, \ref{fig:comp_ind_pca_10nCOM_2}, \ref{fig:comp_ind_pca_15nCYC_2}, and \ref{fig:comp_ind_pca_15nCOM_2}), definitive conclusions should not be drawn. The observed behaviors or patterns in the graphs tend to vary from one problem to another. Therefore, further analysis and investigation are required to fully understand the implications of the PCA results for $9p$ (or more complex) models.

In the pair PCA models, we observed a decreasing trend in variances as the number of parameters increased, namely for the $3p$, $6p$, and $9p$ models. However, it's important to note that the $3p$ models should not be compared in the same manner as the $6p$ and $9p$ models due to the number of PCA components generated. The purpose of the paired graphs ({Figures \ref{fig:comp_pair_pca_4n_CYC}}, {\ref{fig:comp_pair_pca_4n_COM}}, {\ref{fig:comp_pair_pca_10n_CYC}}, {\ref{fig:comp_pair_pca_10n_COM}}, {\ref{fig:comp_pair_pca_15n_CYC}}, {\ref{fig:comp_pair_pca_15n_COM}}) was to determine if the individual behaviors could be captured within a pair PCA. This would suggest that differences between parameters in QAOA models could be detected within the same PCA. In most cases, the individual behaviors were indeed maintained in the pair graphs, supporting the notion that distinct parameter characteristics could be identified using the pair PCA approach.

Moving to the t-SNE analysis, the results presented in {Tables \ref{tab:ind_kl_divergence_t-SNE}}, {\ref{tab:ind_kl_divergence_t-SNE_4nCOM}}, {\ref{tab:ind_kl_divergence_t-SNE_10nCYC}}, {\ref{tab:ind_kl_divergence_t-SNE_10nCOM}}, {\ref{tab:ind_kl_divergence_t-SNE_15nCYC}} and {\ref{tab:ind_kl_divergence_t-SNE_15nCOM}} show a clear tendency of generating better KL-Divergence values for the entangled models, with the difference being more pronounced depending on the perplexity value. In the individual t-SNE analysis, the best results were generally generated by the $3p$ models. This can be attributed to the fact that these models have only 3 parameters, and the t-SNE projection to the plane does not lose a significant amount of information in the process. Additionally, it is important to mention that the best KL values were reported at 99 perplexity, where all the models generated good values that are closer to zero.


The paired t-SNE models presented in Tables \ref{tab:pair_kl_divergence_t-SNE}, \ref{tab:pair_kl_divergence_t-SNE_4nCOM}, \ref{tab:pair_kl_divergence_t-SNE_10nCYC}, \ref{tab:pair_kl_divergence_t-SNE_10nCOM}, \ref{tab:pair_kl_divergence_t-SNE_15nCYC}, and \ref{tab:pair_kl_divergence_t-SNE_15nCOM} also demonstrate that the $3p$ models achieve the best KL values. However, in this case, the quality of the reported values noticeably decreases as the number of parameters increases for the majority of perplexity values tested. Interestingly, the worst KL values were consistently obtained at 30 perplexity, while the best KL values were obtained at 199 perplexity, similar to the individual case at 99 perplexity. Most paired models generated good KL-Divergence values, indicating a better representation in the low-dimensional space.

In the graphical results for the individual t-SNE models shown in Figures \ref{fig:comp_ind_t-SNE_4nCYC}, \ref{fig:comp_ind_t-SNE_4nCOM}, \ref{fig:comp_ind_t-SNE_10nCYC}, \ref{fig:comp_ind_t-SNE_10nCYC_2}, \ref{fig:comp_ind_t-SNE_10nCOM}, \ref{fig:comp_ind_t-SNE_10nCOM_2}, \ref{fig:comp_ind_t-SNE_15nCYC}, \ref{fig:comp_ind_t-SNE_15nCYC_2}, \ref{fig:comp_ind_t-SNE_15nCOM}, and \ref{fig:comp_ind_t-SNE_15nCOM_2}, we selected the graphs generated at 99 perplexity as they offered the best representation based on the KL-Divergence value. In the $3p$ models, we observed similar patterns to those seen in the PCA graphs. The non-entangled $3p$ model (red) consistently exhibited a 3-line cluster pattern across different problems and depths of the QAOA model. On the other hand, the entangled $3p$ model (blue) displayed varying patterns depending on the problem type and QAOA depth, often generating line clustering patterns, although they were not as well-defined as those of the non-entangled model.

For the $6p$ models, we observed similar patterns between the entangled (purple) and non-entangled (green) models. At 99 perplexity, most of the graphs displayed an elliptical pattern, with the entangled models sometimes exhibiting more pronounced grouping behavior in specific areas of the plane. The same behavior observed in the $6p$ models was also reported in the more complex $9p$ models, where at the highest perplexity, both non-entangled (orange) and entangled (brown) models generated elliptical patterns, with the non-entangled models exhibiting a more evenly distributed pattern around the ellipse.

In the analysis of the paired t-SNE graphical results presented in Figures \ref{fig:comp_pair_t-SNE_4nCYC}, \ref{fig:comp_pair_t-SNE_4nCOM}, \ref{fig:comp_pair_t-SNE_10nCYC}, \ref{fig:comp_pair_t-SNE_10nCOM}, \ref{fig:comp_pair_t-SNE_15nCYC}, and \ref{fig:comp_pair_t-SNE_15nCOM}, distinctive patterns and behaviors were observed. In the $3p$ models, clear differences between the non-entangled (red) and entangled (blue) models were noticeable, particularly at certain perplexity values, such as 99 or 199. This suggests that paired t-SNE effectively differentiates between different types of data within the model. Some distributions observed in the individual graphs were preserved in the paired t-SNE plots, while other cases exhibited similarities to the PCA models.

Regarding the $6p$ models, significant disparities were observed between the non-entangled (green) and entangled (purple) models. The non-entangled model displayed a more random distribution in the t-SNE plane across different perplexity values, while the entangled model tended to concentrate in specific areas. At 199 perplexity, both models replicated the elliptical behavior seen in the individual graphs, but the entangled model exhibited clearer distinctions compared to the individual graphs. Specifically, the entangled model demonstrated multiple clusters around a particular distribution, while the non-entangled model continued to exhibit a more evenly distributed pattern.

Similarly, the $9p$ models exhibited behaviors akin to the $6p$ models. The non-entangled model (orange) tended to have a more uniform distribution in the t-SNE plane across various perplexity values, with a distinct elliptical pattern emerging at 199 perplexity. Conversely, the entangled model (brown) showed a propensity to form clusters in specific areas of the plane at different perplexity levels. At 199 perplexity, the entangled models followed the elliptical pattern but appeared more compressed in certain regions of the distribution.


\section{Conclusions}

The application of the PCA method revealed disparities in the variance distribution across PCA components, contingent upon the type of model being tested. Consistently, entangled models exhibited higher variance values, either within each component or in the overall variance.

However, it became apparent that the PCA method was not well-suited for achieving an effective mapping in a low-dimensional space for the investigated problems. As the number of parameters increased, there was a significant reduction in the amount of information captured by the PCA components. Even in the most intricate model tested ($9p$ parameters, $3L$ layers of depth), the first three PCA components usually accounted for less than 50\% of the original variance.

In some instances, the paired PCA models managed to retain the observed patterns from the individual PCA models, which proved valuable in visually distinguishing between entangled and non-entangled models.

In general, t-SNE, whether applied to individual or paired models, outperformed the PCA models. This superiority is evident in the KL-Divergence values obtained at various perplexities, indicating a more effective representation in the low-dimensional space.

In the individual t-SNE models, disparities in KL values between non-entangled and entangled models were also observed. Typically, the entangled models displayed better KL values, which can be attributed to the entanglement stage in the mixing operator. This stage augments the amount of information regarding the relationships between parameter rotations, which can be detected by the t-SNE models.

Lastly, in the paired t-SNE models, notable distinctions were noted between non-entangled and entangled models at different perplexity values. For $3p$ models, linear pattern distributions were more prevalent, while $6p$ and $9p$ models exhibited more random and elliptical distributions for non-entangled models, whereas entangled models showed a proclivity for clustering and adhering to certain patterns. These findings underscore the capacity of t-SNE to visually discriminate between the data relationships of non-entangled and entangled models.


In order to advance future research, it is essential to delve further into the interpretation of the observed distributions. It is currently premature to draw conclusions about the universality of these specific patterns in all models featuring an entanglement stage, regardless of the problem. Moreover, it remains uncertain whether different patterns may arise in other problem types, indicating the presence or absence of an entanglement stage in QAOA. A comprehensive understanding of these phenomena necessitates additional exploration and analysis.

Furthermore, it would be valuable to investigate alternative optimization methods for QAOA to compare the obtained results. Such an analysis would help identify which behaviors persist across different optimization methods, and which ones are influenced by the specific approach employed to solve the presented problems.

\section{Acknowledgements}


The author, B. Sarmina, would like to express sincere gratitude and acknowledge the support of CONACYT (National Council for Science and Technology) of Mexico for providing a scholarship during his Ph.D. program in computer science at the Center for Computing Research of Instituto Politécnico Nacional. 

The authors also acknowledge the support of the Secretaría de Investigación y Posgrado (SIP) from Instituto Politécnico Nacional through the SIP grants: SIP 20220355, SIP 20220865, and SIP 20230316.


\bibliographystyle{plain}

\onecolumn\newpage
\appendix

\section{PCA variances and graphs}

In this appendix, we provide additional results that complement the experiments developed to enhance the breadth of insights and conclusions drawn from the PCA analysis applied to the QAOA-based solutions of max-cut problems. These supplementary findings offer a more comprehensive understanding of the variances obtained and contribute to a more robust evaluation of the overall outcomes.

\begin{table}[ht]
\centering
\begin{tabular}{|c|c|c|c|}
\hline
Parameters & PCA 1      & PCA 2      & PCA 3  \\ \hline
3 parameters         & 0.50298884 & 0.311534 & 0.18547716 \\ \hline
3 parameters ent     & 0.57473804 & 0.42450533 & 0.00075663 \\ \hline
6 parameters         & 0.2283193 & 0.21189936 & 0.18558982 \\ \hline
6 parameters ent     & 0.23935453 & 0.22027888 & 0.18451389 \\ \hline
\end{tabular}
\caption{Individual PCA projections explained variance (4n complete) for the first 3 PCA components.}
\label{tab:exp_var_indPCA_4nCOM}
\end{table}

The table displaying the PCA explained variances for the 4n complete max-cut problem can be found in {Table \ref{tab:exp_var_indPCA_4nCOM}}. Upon analyzing the $3p$ models, it becomes evident that the entangled model exhibits higher variances in PCA 1 and PCA 2. Notably, the entangled model displays an intriguing characteristic where the variance in PCA 3 is nearly zero. This phenomenon is entirely distinct from the previous observed variances in the 4n cyclic problem. When examining the $6p$ models, their variances align more closely with those observed in the previous problem. Moreover, in both entangled models, the variances in PCA 1 demonstrate an increase when compared to the same number of parameter models lacking entanglement.

\begin{figure}[ht]
\centering
\includegraphics[width=10cm, height=8cm]{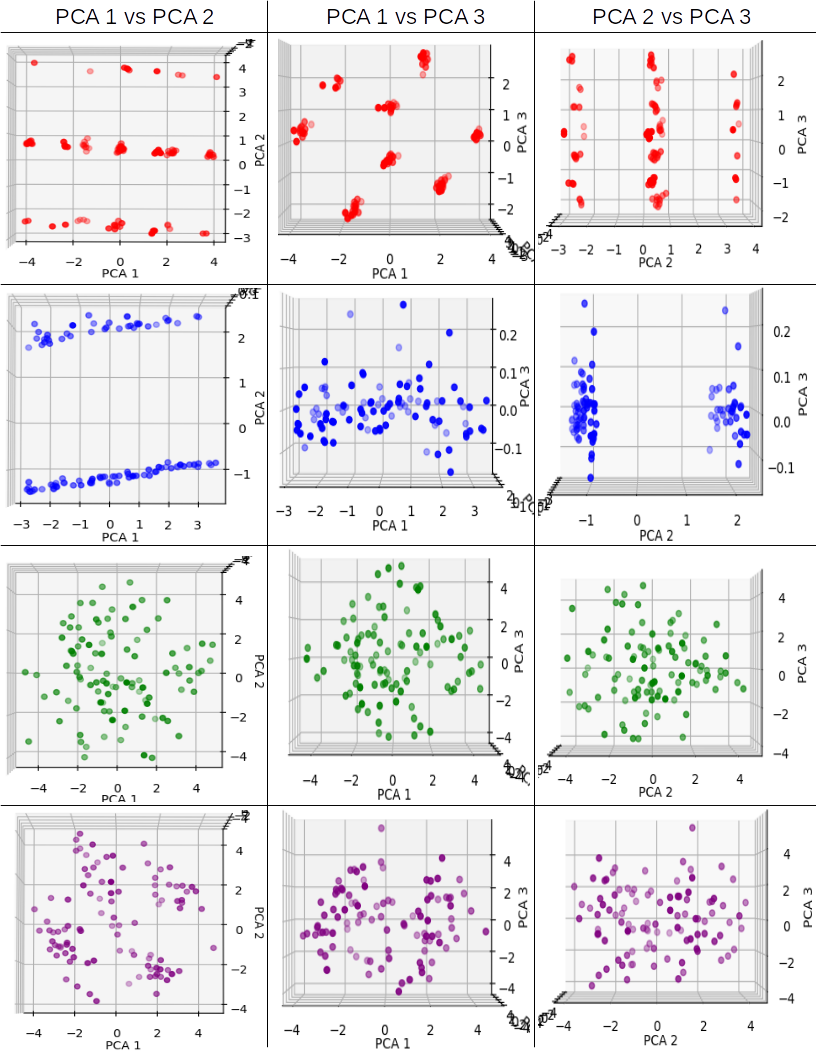}
\caption{PCA individual graphs for 4n complete configuration max-cut problem solved using QAOA, first 3 components. Red corresponds to the $3p$ parameter $1L$ non-entangled, blue $3p$ parameter $1L$ entangled, green $6p$ parameter $2L$ non-entangled and purple $6p$ parameter $2L$ entangled model.}
\label{fig:comp_ind_pca_4nCOM}
\end{figure}

Upon examining {Figure \ref{fig:comp_ind_pca_4nCOM}}, several noteworthy observations come to light. In the case of the $3p$ models, the behavior of the non-entangled model (depicted in red) closely resembles that of the previous problem. The most prominent change occurs in the PCA 1 vs PCA 2 graph, where a similar linear distribution is observed but with a distinct orientation. Conversely, the entangled model (depicted in blue) exhibits a novel pattern characterized by two distinct clusters. It is intriguing to note that these two clusters manifest in both the PCA 1 vs PCA 2 and PCA 2 vs PCA 3 graphs, albeit from different perspectives. Moving on to the $6p$ models, the green model (non-entangled) displays a distribution similar to the previous problem, featuring a random dispersion of points across different plane perspectives, devoid of distinguishable clusters or patterns. However, the entangled model (depicted in purple) reveals the recurrence of the three-cluster behavior observed in the cyclic problem, specifically in the PCA 1 vs PCA 2 graph. This phenomenon potentially signifies an increase in data correlations when the entanglement stage is implemented.

\begin{table}[ht!]
\centering
\begin{tabular}{|c|c|c|c|}
\hline
Parameters & PCA 1      & PCA 2      & PCA 3 \\ \hline
3 parameters         & 0.4655341 & 0.34885069 & 0.18561521 \\ \hline
6 parameters         & 0.21851466 & 0.18343424 & 0.17652118 \\ \hline
\end{tabular}
\caption{Pair PCA projections explained variance for the first 3 PCA components for the 4n complete max-cut problem.}
\label{tab:exp_var_pairPCA_4nCOM}
\end{table}

The presented explained variances for the pair PCA models, representing the $3p$ and $6p$ (equivalent to $1L$ and $2L$ of depth, respectively), in {Table \ref{tab:exp_var_pairPCA_4nCOM}} pertain to the 4n complete max-cut problem. These variances exhibit similarities to those observed in the cyclic problem, with the main distinction being a notable decrease in the PCA 3 component for the $3p$ pair PCA compared to the previous problem. For the $6p$ pair PCA, the variances remain relatively consistent, with differences of at most $0.3$ in the first three PCA components. Overall, the variances obtained in the pair PCA analysis for the 4n complete max-cut problem are largely comparable to those observed in the cyclic problem, with only slight variations in specific components.

\begin{figure}[ht]
\centering
\includegraphics[width=12cm, height=8cm]{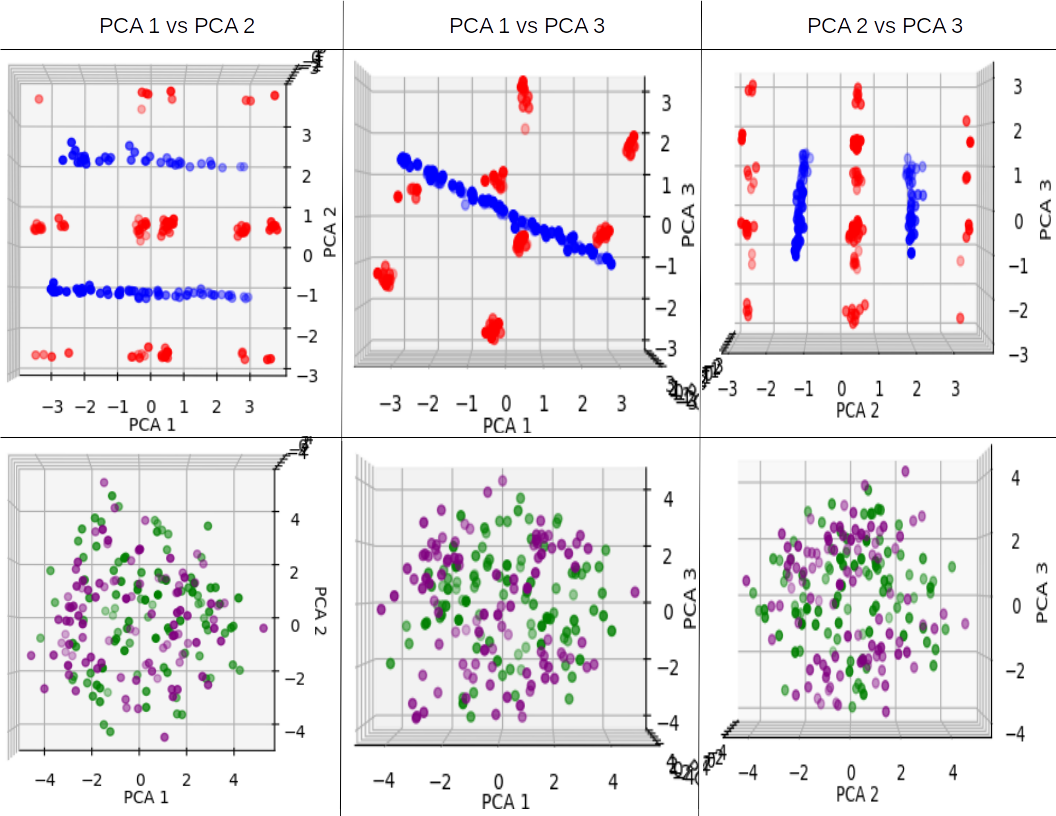}
\caption{PCA pair graphs for 4n complete configuration max-cut problem solved using QAOA, first 3 components. Red corresponds to the $3p$ parameter $1L$ non-entangled, blue $3p$ parameter $1L$ entangled, green $6p$ parameter $2L$ non-entangled and purple $6p$ parameter $2L$ entangled model.}
\label{fig:comp_pair_pca_4n_COM}
\end{figure}

The pair PCA graphs illustrating the 4n complete max-cut configuration problem can be observed in {Figure \ref{fig:comp_pair_pca_4n_COM}}. For the $3p$ models (represented in red and blue), the PCA 1 vs PCA 2 graph exhibits patterns akin to those observed in the individual PCA graphs. In the PCA 2 vs PCA 3 graph, the pattern appears to be preserved for the non-entangled model, while the entangled model displays a distribution of points that are closer together. Notably, the blue points in the PCA 1 vs PCA 3 graph adhere to a distinct line pattern that clearly differentiates the projection of the non-entangled and entangled models. Shifting focus to the $6p$ models (depicted in green and purple), the first PCA 1 vs PCA 2 graph depicts a scattered distribution of points for the non-entangled model (green), while the entangled model (purple) exhibits two lightly clustered areas that are barely discernible. However, in the PCA 1 vs PCA 3 graph, a conspicuous pattern of three elliptic clusters emerges exclusively in the entangled model. Furthermore, when considering the PCA 2 and PCA 3 graph, which showcases two clusters within the entangled model, these clusters can be interpreted as the pair PCA model's ability to detect specific correlations between the non-entangled and entangled data due to the diverse distribution of values from the different models. Overall, the pair PCA graphs provide insights indicating that the entanglement stage can unveil additional information regarding the correlations between different models.

\begin{table}[ht!]
\centering
\begin{tabular}{|c|c|c|c|}
\hline
Parameters & PCA 1      & PCA 2      & PCA 3  \\ \hline
3 parameters         & 0.4824018 & 0.32759113 & 0.19000707 \\ \hline
3 parameters ent     & 0.42582802 & 0.29568193 & 0.27849004 \\ \hline
6 parameters         & 0.22421832 & 0.19958267 & 0.19614923 \\ \hline
6 parameters ent     & 0.27777425 & 0.20329209 & 0.18833349 \\ \hline
9 parameters         & 0.1681149 & 0.14530348 & 0.13463924 \\ \hline
9 parameters ent     & 0.17743445 & 0.1615457 & 0.14796169 \\ \hline
\end{tabular}
\caption{Individual PCA projections explained variance (10n cyclic) for the first 3 PCA components.}
\label{tab:exp_var_indPCA_10nCYC}
\end{table}

The explained variances for the 10n cyclic max-cut problem can be found in {Table \ref{tab:exp_var_indPCA_10nCYC}}. In this particular problem, we have collected results for three different levels of depth: $1L$ (corresponding to $3p$ parameters), $2L$ (equivalent to $6p$ parameters), and $3L$ (comprising $9p$ parameters). The first two models, $1L$ and $2L$, exhibit similar results to the previous cyclic problem involving 4n. However, it is crucial to note that the entangled models with $6p$ and $9p$ parameters demonstrate an increase in variance within the PCA 1 component compared to the non-entangled models. Moreover, the PCA 2 and PCA 3 components display a higher variance in general compared to the non-entangled models, including the $3p$ model. This rise in variance around the components can be attributed to the introduction of the entanglement stage, which amplifies the covariances between the elements.

\begin{figure}[ht]
\centering
\includegraphics[width=10cm, height=8cm]{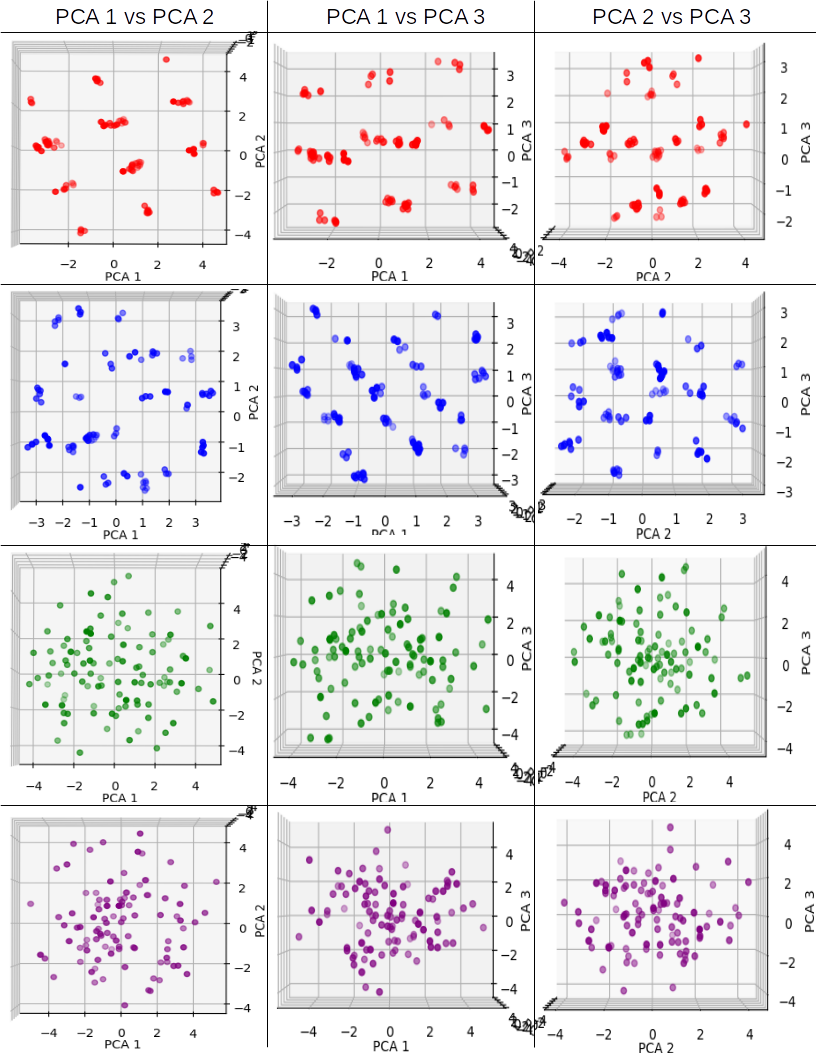}
\caption{PCA individual graphs for 10n cyclic configuration max-cut problem solved using QAOA, first 3 components. Red corresponds to the $3p$ parameter $1L$ non-entangled, blue $3p$ parameter $1L$ entangled, green $6p$ parameter $2L$ non-entangled and purple $6p$ parameter $2L$ entangled model.}
\label{fig:comp_ind_pca_10nCYC}
\end{figure}

The graphs illustrating the $3p$ and $6p$ parameter models for the 10n cyclic max-cut problem can be observed in {Figure \ref{fig:comp_ind_pca_10nCYC}}. In the case of the first $3p$ non-entangled model (depicted in red), the distribution closely resembles that of the first cyclic problem. However, for the entangled model (depicted in blue), the data projection exhibits a distinct pattern that is not readily recognizable. Moving on to the $6p$ parameter models, there are some similarities in behavior compared to the previous cyclic problem. The non-entangled model (depicted in green) displays a random distribution of points without any discernible clusters or patterns. Conversely, in the entangled model (depicted in purple), the PCA 1 vs PCA 2 graph reveals a prominent cluster in the center, accompanied by two smaller clusters on either side. This distribution bears resemblance to the pattern observed in the previous problem. Furthermore, in the PCA 1 vs PCA 3 and PCA 2 vs PCA 3 graphs of the entangled model (purple), the points appear to converge towards the center of each graph, reinforcing the clustering effect. Overall, these observations suggest that while the non-entangled models maintain similar distributions to the previous problem, the entangled models exhibit distinct patterns that may indicate the impact of the entanglement stage on the data projection.

\begin{figure}[ht]
\centering
\includegraphics[width=12cm, height=8cm]{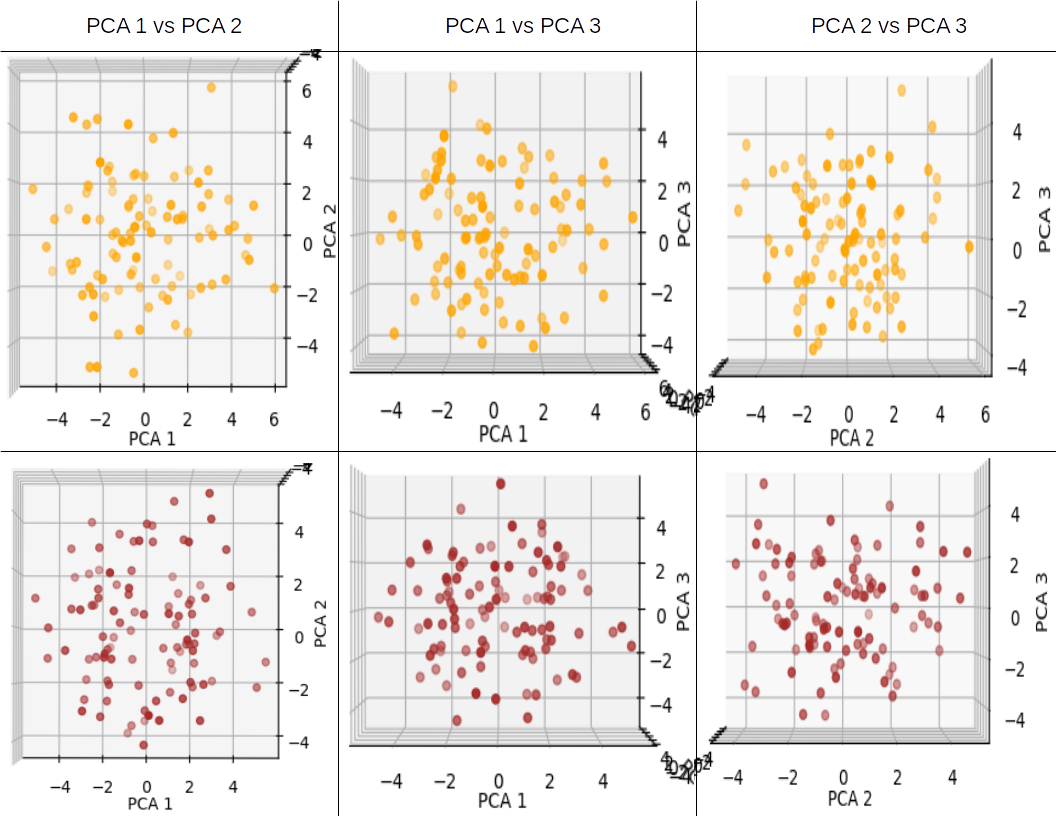}
\caption{PCA individual graphs for 10n cyclic configuration max-cut problem solved using QAOA, first 3 components. Orange $9p$ parameter $3L$ non-entangled and purple $9p$ parameter $3L$ entangled model.}
\label{fig:comp_ind_pca_10nCYC_2}
\end{figure}

In the last $3L$ depth model with $9p$ parameters ({Figure \ref{fig:comp_ind_pca_10nCYC_2}}, the distribution for the non-entangled model (orange) is very similar to the $6p$ non-entangled model, with a random distribution of points and no distinguishable clusters. For the entangled model, only the PCA 1 vs PCA 2 graph seems to have a pattern, with two light clusters divided by a central line.

\begin{table}[ht!]
\centering
\begin{tabular}{|c|c|c|c|}
\hline
Parameters & PCA 1      & PCA 2      & PCA 3 \\ \hline
3 parameters         & 0.45436643 & 0.31238244 & 0.23325113 \\ \hline
6 parameters         & 0.21821814 & 0.1962479 & 0.17828477 \\ \hline
9 parameters         & 0.14672728 & 0.13850982 & 0.12026834 \\ \hline
\end{tabular}
\caption{Pair PCA projections explained variance for the first 3 PCA components for the 10n cyclic max-cut problem.}
\label{tab:exp_var_pairPCA_10nCYC}
\end{table}


The pair PCA explained variances for the 10n cyclic max-cut problem are presented in {Table \ref{tab:exp_var_indPCA_10nCYC}}. The variances for the $3p$ and $6p$ models were very similar to the previous results, with a considerable decrease in the amount of variance represented in each PCA component as the number of parameters increased. This trend persists with the $9p$ pair PCA model values.

\begin{figure}[ht]
\centering
\includegraphics[width=10cm, height=8cm]{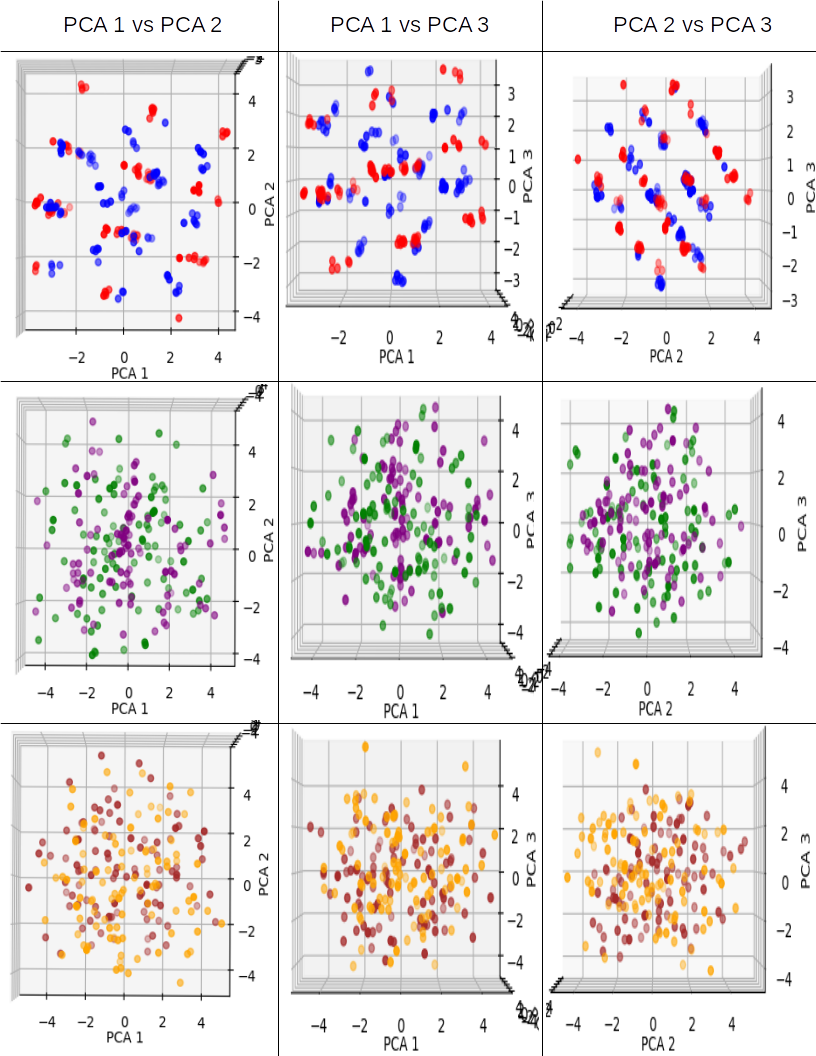}
\caption{PCA pair graphs for 10n cyclic configuration max-cut problem solved using QAOA, first 3 components. Red corresponds to the $3p$ parameter $1L$ non-entangled, blue $3p$ parameter $1L$ entangled, green $6p$ parameter $2L$ non-entangled, purple $6p$ parameter $2L$ entangled model, orange $9p$ parameter $3L$ non-entangled and brown $9p$ parameter $3L$ entangled model.}
\label{fig:comp_pair_pca_10n_CYC}
\end{figure}

The pair PCA graphs are presented in {Figure \ref{fig:comp_pair_pca_10n_CYC}}. The $3p$ and $6p$ models have similar behaviors to the previous cyclic problem. The $3p$ model preserves almost the same distribution of points in the projection as the individual graphs, while for the $6p$ models, the green (non-entangled) model has a random distribution of points similar to the individual graph. However, the purple ($6p$ entangled) model exhibits a clear cluster pattern behavior for the PCA 1 vs PCA 2 and PCA 1 vs PCA 3 graphs. In the more complex models with $9p$, there is no clear behavior of the projection distribution, and due to the low variance for each PCA component presented in the graph, we cannot establish a precise interpretation of the results because the PCA mapping has lost a lot of original information.

\begin{table}[ht!]
\centering
\begin{tabular}{|c|c|c|c|}
\hline
Parameters & PCA 1      & PCA 2      & PCA 3  \\ \hline
3 parameters         & 0.49767238 & 0.33234628 & 0.16998134 \\ \hline
3 parameters ent     & 0.588512290 & 0.411087639 & 0.0004000709 \\ \hline
6 parameters         & 0.24720666 & 0.19582826 & 0.16085243 \\ \hline
6 parameters ent     & 0.23681691 & 0.19549636 & 0.17376718 \\ \hline
9 parameters         & 0.17181209 & 0.14208608 & 0.12933972 \\ \hline
9 parameters ent     & 0.15950657 & 0.14702026 & 0.14211113 \\ \hline
\end{tabular}
\caption{Individual PCA projections explained variance (10n complete) for the first 3 PCA components.}
\label{tab:exp_var_indPCA_10nCOM}
\end{table}

The explained variances for the first 3 PCA components in the individual models for the 10n complete problem are presented in {Table \ref{tab:exp_var_indPCA_10nCOM}}. Comparing the table with the individual variances for the cyclic problem, we observe some interesting results. Starting with the $3p$ model, the entangled version shows an increase in the amount of variance associated with PCA 1 and PCA 2 compared to the non-entangled model, which is the opposite of what was observed in the cyclic problem. For the $6p$ and $9p$ models, the entangled versions show a decrease in the amount of variance contained in PCA 1 and PCA 2 components with respect to the non-entangled models, while PCA 3 has a greater value in general. Analyzing these results with the cyclic problem, we can observe how the problem's structure modifies how the entanglement stage in the mixing operator can affect the variance distribution along the PCA components. However, the difference between the components in the entangled models seems to be lower compared to the non-entangled ones. Another important observation is that in the $6p$ and $9p$ models, the total amount of variance captured by the first 3 PCA components is slightly higher in the entangled models compared to the non-entangled ones. 

\begin{figure}[ht]
\centering
\includegraphics[width=10cm, height=8cm]{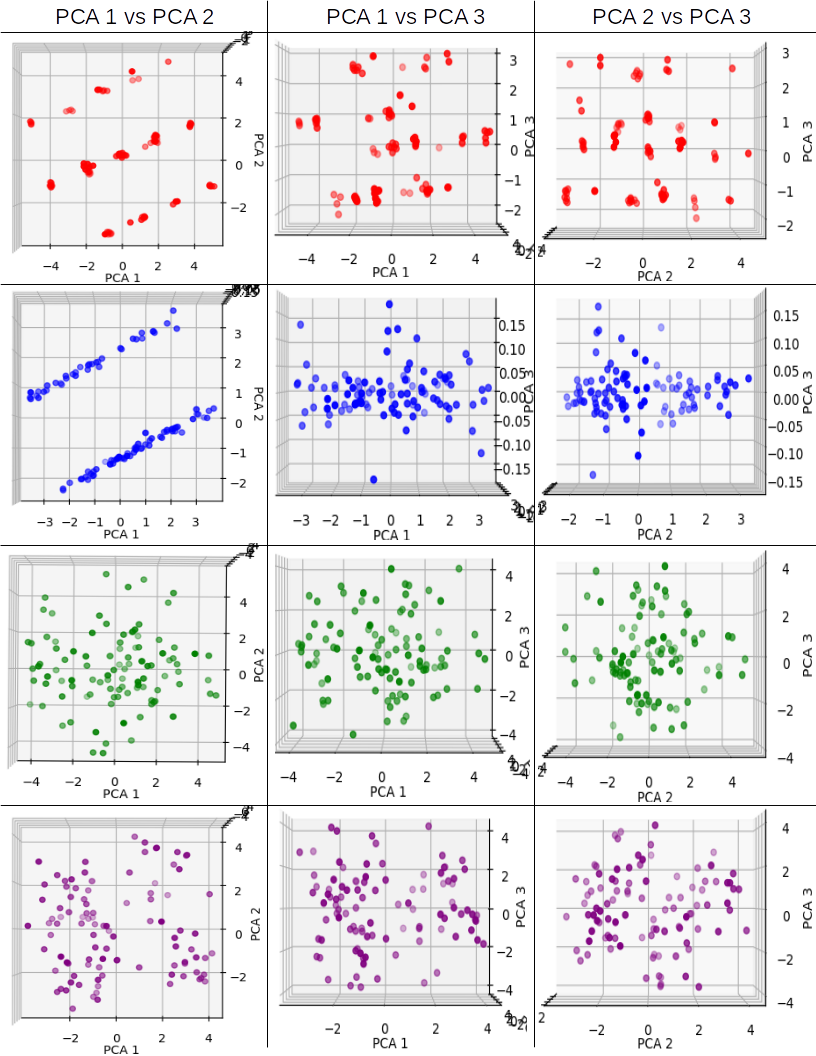}
\caption{PCA individual graphs for 10n complete configuration max-cut problem solved using QAOA, first 3 components. Red corresponds to the $3p$ parameter $1L$ non-entangled, blue $3p$ parameter $1L$ entangled, green $6p$ parameter $2L$ non-entangled and purple $6p$ parameter $2L$ entangled model.}
\label{fig:comp_ind_pca_10nCOM}
\end{figure}

The individual graphs presented in {Figure \ref{fig:comp_ind_pca_10nCOM}} exhibit a behavior similar to that of the 4n complete max-cut problem. The $3p$ non-entangled (red) and entangled (blue) models have almost the same type of distribution for the PCA 1 vs PCA 2, albeit with different orientations. In the case of the $6p$ non-entangled model (green), it has a similar random distribution as in the previous problems (not only the complete problems). Meanwhile, the $6p$ entangled model (purple) has a major cluster on the left and two small clusters on the right in the PCA 1 vs PCA 2 graph, and, the PCA 2 vs PCA 3 graph has a two-cluster distribution.

\begin{figure}[ht]
\centering
\includegraphics[width=12cm, height=8cm]{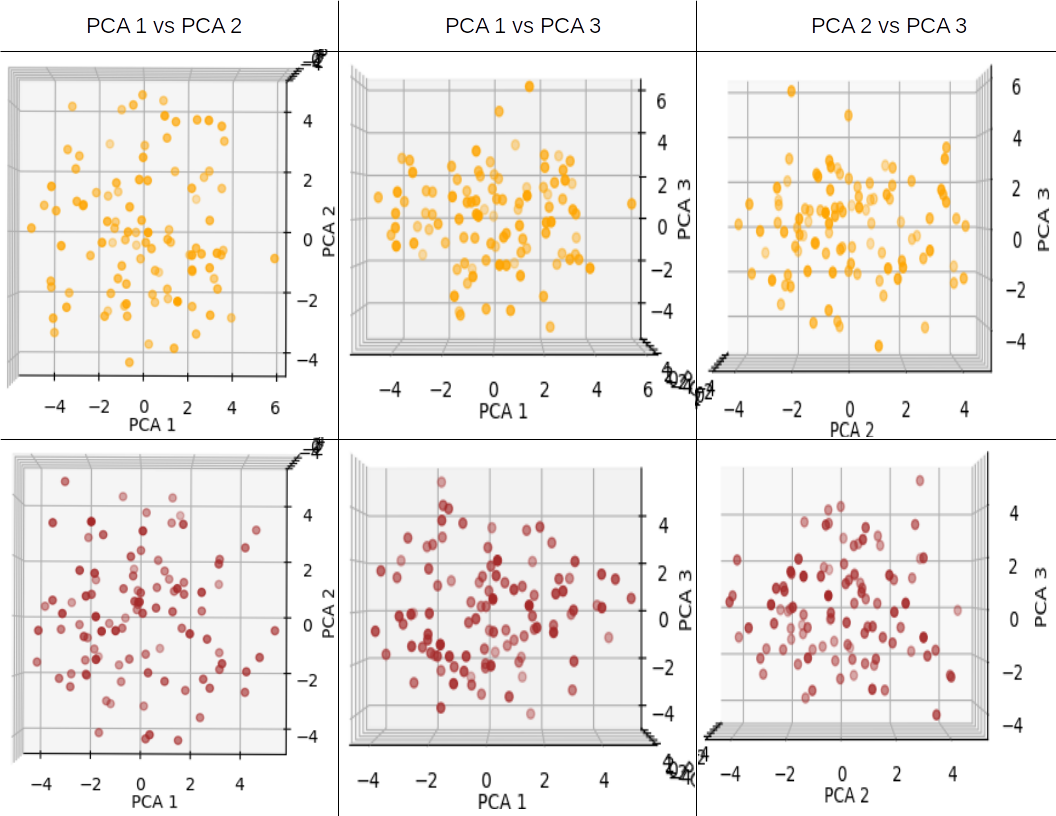}
\caption{PCA individual graphs for 10n complete configuration max-cut problem solved using QAOA, first 3 components. Orange $9p$ parameter $3L$ non-entangled and brown $9p$ parameter $3L$ entangled model.}
\label{fig:comp_ind_pca_10nCOM_2}
\end{figure}

Completing the individual graphs in the 10n complete max-cut problems, we have the $9p$ parameter models ($3L$ depth) in {Figure \ref{fig:comp_ind_pca_10nCOM_2}}. In this case, there is no clear behavior in the non-entangled and entangled models. This is not surprising because the amount of variance that each perspective has is very low, and it cannot give a correct representation of the original information.

\begin{table}[ht!]
\centering
\begin{tabular}{|c|c|c|c|}
\hline
Parameters & PCA 1      & PCA 2      & PCA 3 \\ \hline
3 parameters         & 0.46514143 & 0.34410486 & 0.19075371 \\ \hline
6 parameters         & 0.20148807 & 0.19106115 & 0.17569887 \\ \hline
9 parameters         & 0.1569094 & 0.137659 & 0.12379254 \\ \hline
\end{tabular}
\caption{Pair PCA projections explained variance for the first 3 PCA components for the 10n complete max-cut problem.}
\label{tab:exp_var_pairPCA_10nCOM}
\end{table}

The explained variances for the 10n complete max-cut problem are presented in {Table \ref{tab:exp_var_indPCA_10nCOM}}. The values of variance are very similar to the ones obtained in the cyclic problem, showing a decreasing trend in importance (due to the decrease in variance) with an increase in the number of parameters. In the case of the $9p$ models, the amount of variance in the first 3 components is less than $40\%$.

\begin{figure}[ht]
\centering
\includegraphics[width=10cm, height=8cm]{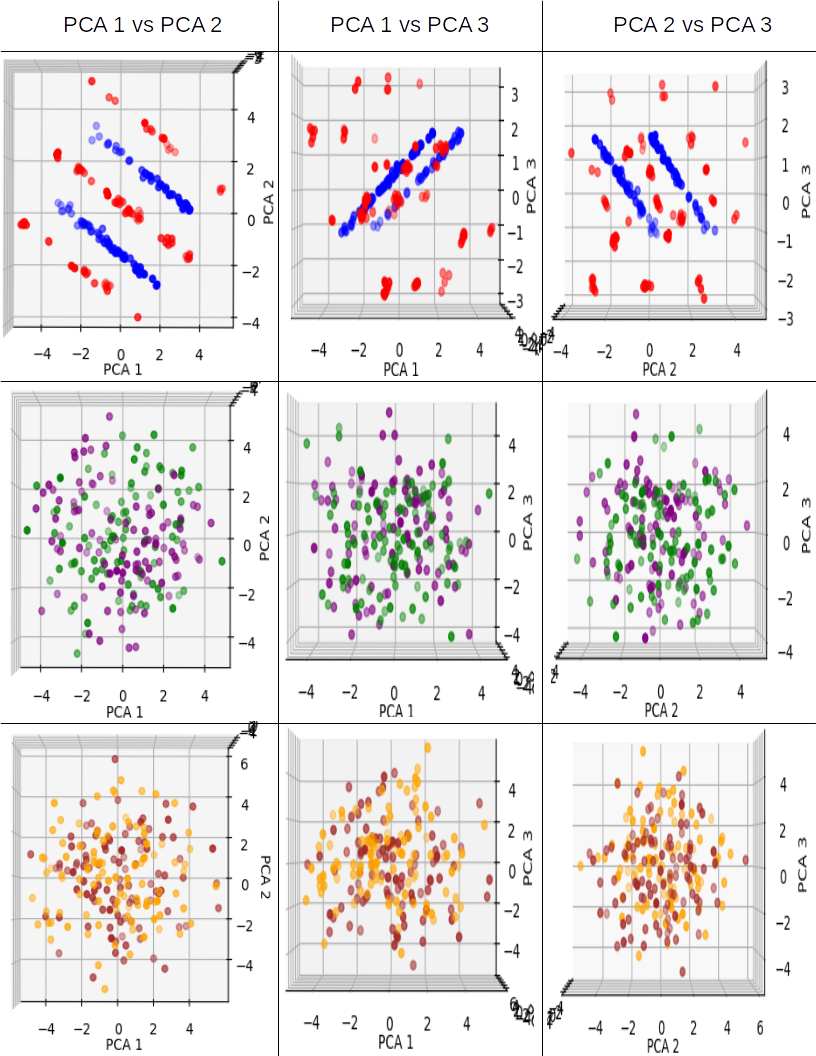}
\caption{PCA pair graphs for 10n complete configuration max-cut problem solved using QAOA, first 3 components. Red corresponds to the $3p$ parameter $1L$ non-entangled, blue $3p$ parameter $1L$ entangled, green $6p$ parameter $2L$ non-entangled, purple $6p$ parameter $2L$ entangled model, orange $9p$ parameter $3L$ non-entangled and brown $9p$ parameter $3L$ entangled model.}
\label{fig:comp_pair_pca_10n_COM}
\end{figure}

The pair PCA model for the 10n complete max-cut problem is presented in Figure \ref{fig:comp_pair_pca_10n_COM}. In the case of the $3p$ models, the distribution is very similar to the individual graphs, with changes observed in the PCA 1 vs PCA 3 and PCA 2 vs PCA 3 distributions for the entangled model (blue). For the $6p$ parameter models, the distribution shows no clear pattern or clusters, with only two light clusters and some outliers in the PCA 1 vs PCA 2 plot. However, due to the low variance of the PCA components, these results cannot be considered conclusive. Finally, for the $9p$ models, the distribution appears to be random, with no clear patterns observed. Again, due to the low variance, these results are to be expected.

\begin{table}[ht!]
\centering
\begin{tabular}{|c|c|c|c|}
\hline
Parameters & PCA 1      & PCA 2      & PCA 3  \\ \hline
3 parameters         & 0.4647624 & 0.32860527 & 0.20663233 \\ \hline
3 parameters ent     & 0.38826281 & 0.34828084 & 0.26345635 \\ \hline
6 parameters         & 0.22713273 & 0.19519285 & 0.18337327 \\ \hline
6 parameters ent     & 0.27183177 & 0.23569836 & 0.19037519 \\ \hline
9 parameters         & 0.1592166 & 0.1491276 & 0.14043665 \\ \hline
9 parameters ent     & 0.17716564 & 0.15707367 & 0.13968417 \\ \hline
\end{tabular}
\caption{Individual PCA projections explained variance (15n cyclic) for the first 3 PCA components.}
\label{tab:exp_var_indPCA_15nCYC}
\end{table}

The results shown in Table \ref{tab:exp_var_indPCA_15nCYC} exhibit similar behavior to those observed in the 10n cyclic problem. For the $3p$ models, the non-entangled model demonstrates greater values for PCA 1 and PCA 2, whereas the entangled model produces a more evenly distributed variance in PCA 2 and PCA 3. For the $6p$ models, the entangled model displays higher variance values for PCA 1 and PCA 2, as well as for the first three components, which is consistent with the earlier findings. In the case of the $9p$ models, the entangled model has higher values for all PCA components, although the difference is not substantial. Overall, these results suggest that the entangled models generally perform better in terms of the amount of variance information that the model is able to detect and project in the new PCA space.

\begin{figure}[ht]
\centering
\includegraphics[width=10cm, height=8cm]{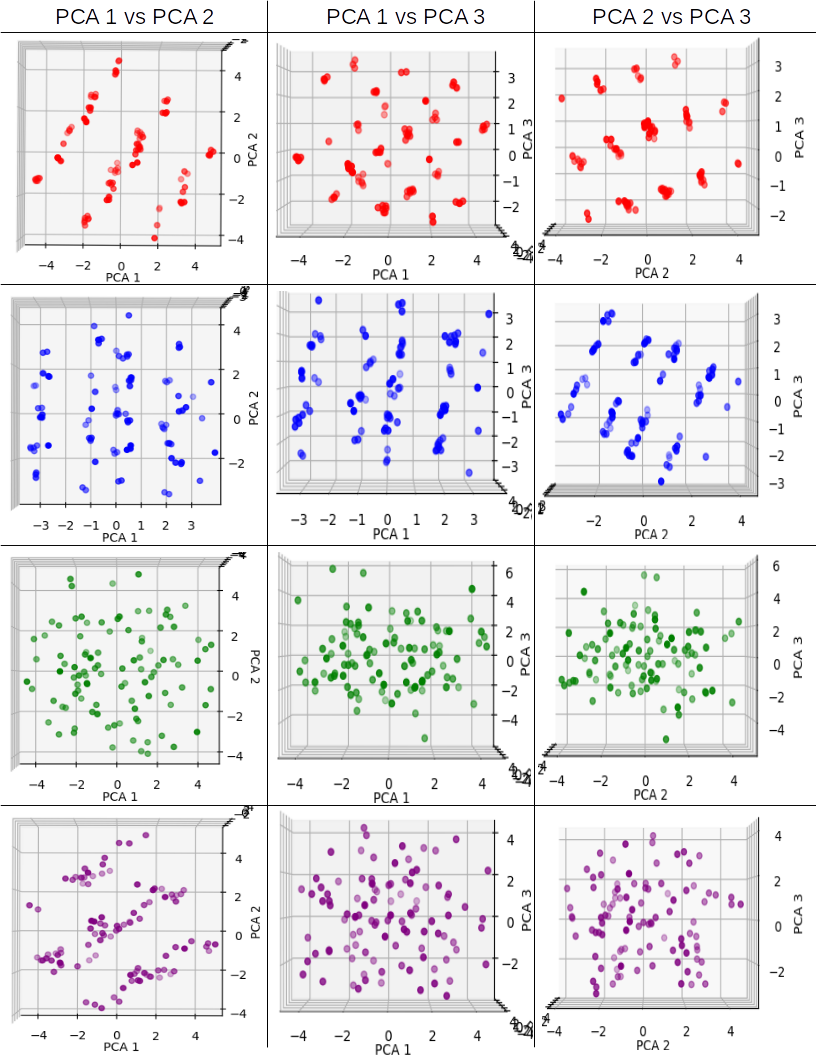}
\caption{PCA individual graphs for 15n cyclic configuration max-cut problem solved using QAOA, first 3 components. Red corresponds to the $3p$ parameter $1L$ non-entangled, blue $3p$ parameter $1L$ entangled, green $6p$ parameter $2L$ non-entangled and purple $6p$ parameter $2L$ entangled model.}
\label{fig:comp_ind_pca_15nCYC}
\end{figure}

For the individual PCA graphs of the 15n cyclic max-cut problem, refer to {Figure \ref{fig:comp_ind_pca_15nCYC}}. In the $3p$ models, both non-entangled (red) and entangled (blue), we observe a behavior similar to previous experiments. Particularly, interesting patterns can be observed in the PCA 1 vs PCA 2 and PCA 2 vs PCA 3 planes. Shifting our focus to the $6p$ models, the non-entangled model (green) exhibits patterns consistent with previous observations, with no clear discernible behavior or pattern across different PCA planes. However, for the entangled model (purple), the presence of the three-line clustering behavior, previously observed in the PCA 1 vs PCA 2 plane for the 4n and 10n cyclic problems, reappears.

\begin{figure}[ht]
\centering
\includegraphics[width=12cm, height=8cm]{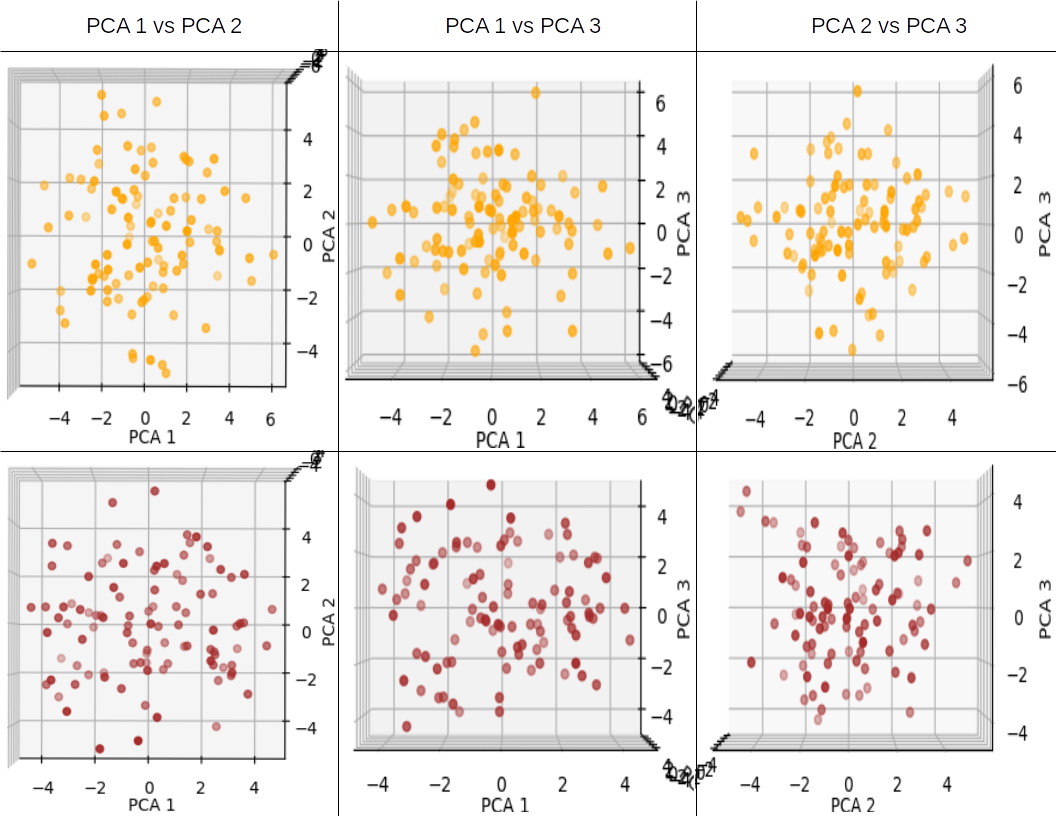}
\caption{PCA individual graphs for 15n cyclic configuration max-cut problem solved using QAOA, first 3 components. Orange $9p$ parameter $3L$ not entangled and brown $9p$ parameter $3L$ entangled model.}
\label{fig:comp_ind_pca_15nCYC_2}
\end{figure}

In the $9p$ parameters models ({Figure \ref{fig:comp_ind_pca_15nCYC_2}}), no clear patterns can be distinguished in both the non-entangled (yellow) and entangled (brown) models. This lack of clear patterns is not surprising, considering the low variance associated with the first 3 PCA components. As previously mentioned, when the variance is low, it becomes more challenging to achieve a meaningful mapping in the low-dimensional space using PCA.

\begin{table}[ht!]
\centering
\begin{tabular}{|c|c|c|c|}
\hline
Parameters & PCA 1      & PCA 2      & PCA 3 \\ \hline
3 parameters         & 0.41927731 & 0.34498266 & 0.23574004 \\ \hline
6 parameters         & 0.23265555 & 0.18633541 & 0.18280923 \\ \hline
9 parameters         & 0.15291139 & 0.14312175 & 0.12984339 \\ \hline
\end{tabular}
\caption{Pair PCA projections explained variance for the first 3 PCA components for the 15n cyclic max-cut problem.}
\label{tab:exp_var_pairPCA_15nCYC}
\end{table}

Now, regarding the pair PCA models in the 15n cyclic problem, we observe similar behavior as in the previous cyclic problem. The $3p$ models present the best PCA values, which is not surprising since this model has the same dimension as the PCA components. The $6p$ models accumulate approximately 60\% of the variance in the original data for the first 3 PCA components, making it the second-best performing model. Finally, the $9p$ models have lower PCA values, with less than 40\% of the variance of the data in the first 3 components.

\begin{figure}[ht]
\centering
\includegraphics[width=10cm, height=8cm]{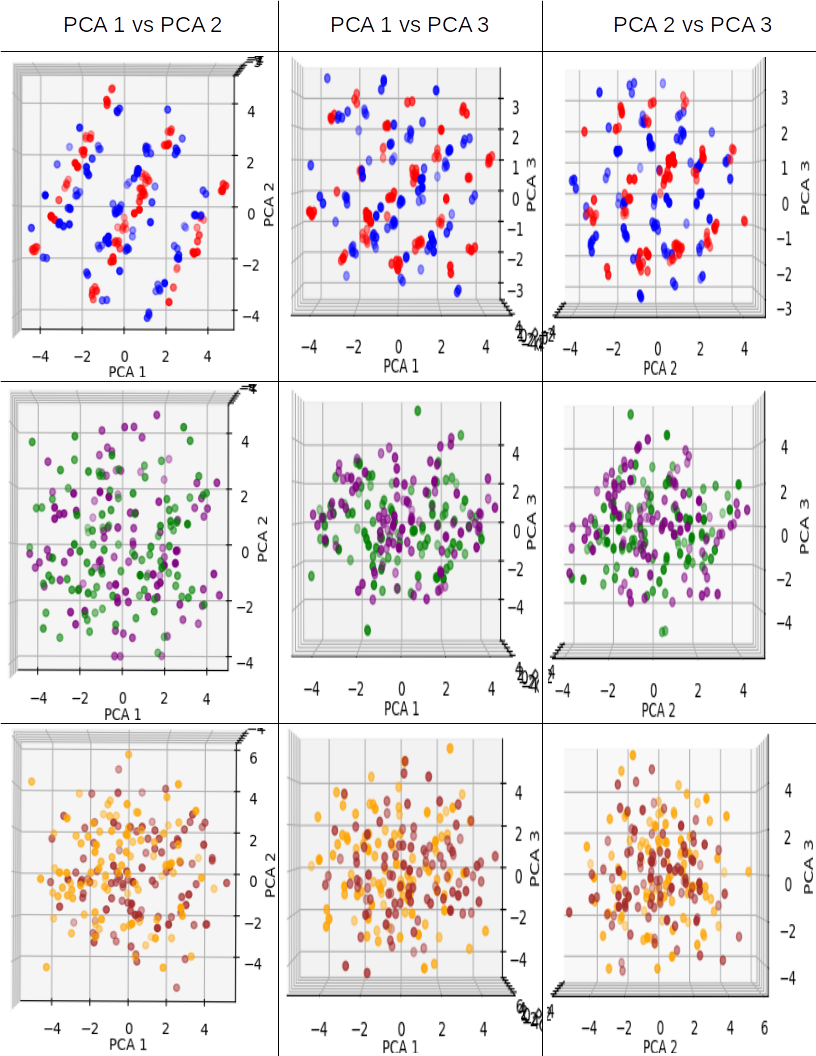}
\caption{PCA pair graphs for 15n cyclic configuration max-cut problem solved using QAOA, first 3 components. Red corresponds to the $3p$ parameter $1L$ non-entangled, blue $3p$ parameter $1L$ entangled, green $6p$ parameter $2L$ non-entangled, purple $6p$ parameter $2L$ entangled model, orange $9p$ parameter $3L$ non-entangled and brown $9p$ parameter $3L$ entangled model.}
\label{fig:comp_pair_pca_15n_CYC}
\end{figure}

The results presented in {Figure \ref{fig:comp_pair_pca_15n_CYC}} exhibit similar patterns to those observed in previous problems for the $3p$ and $6p$ models. In particular, for the $6p$ models, both the entangled (purple) and non-entangled (green) models continue to exhibit their respective distribution behaviors. The non-entangled model displays a scattered distribution in the PCA 1 vs PCA 2 plane, while the entangled model demonstrates clustering behavior in the PCA 1 vs PCA 3 plane. However, for the $9p$ models, neither the entangled (brown) nor the non-entangled (yellow) models exhibit clear patterns. The only noticeable difference is that the data points in the entangled model tend to be closer together, although this distinction is difficult to discern.

\begin{table}[ht!]
\centering
\begin{tabular}{|c|c|c|c|}
\hline
Parameters & PCA 1      & PCA 2      & PCA 3  \\ \hline
3 parameters         & 0.41775729 & 0.35261959 & 0.22962311 \\ \hline
3 parameters ent     & 0.38723863 & 0.36655932 & 0.24620205 \\ \hline
6 parameters         & 0.22555648 & 0.19689321 & 0.18005628 \\ \hline
6 parameters ent     & 0.28402724 & 0.21446195 & 0.18652736 \\ \hline
9 parameters         & 0.16446156 & 0.15145779 & 0.14506611 \\ \hline
9 parameters ent     & 0.18999579 & 0.16075822 & 0.13010846 \\ \hline
\end{tabular}
\caption{Individual PCA projections explained variance (15n complete) for the first 3 PCA components.}
\label{tab:exp_var_indPCA_15nCOM}
\end{table}

The final problem examined using PCA is the 15n complete configuration max-cut problem. The individual PCA variances are presented in {Table \ref{tab:exp_var_indPCA_15nCOM}}. In the $3p$ models (both entangled and non-entangled), the behavior aligns with the previous findings. However, in the $6p$ models, the distribution of explained variance differs from the 10n complete configuration problem. Here, the entangled model demonstrates a noticeable increase in variance due to the presence of the entanglement stage, resembling the behavior observed in the cyclic problems. Similarly, in the $9p$ models, the entangled QAOA exhibits a higher total amount of variance in the first 3 PCA components, mirroring the results observed in the $6p$ models. Additionally, consistent with the 10n problem, the total amount of variance is higher in the entangled models for both the $6p$ and $9p$ cases.

\begin{figure}[ht]
\centering
\includegraphics[width=10cm, height=8cm]{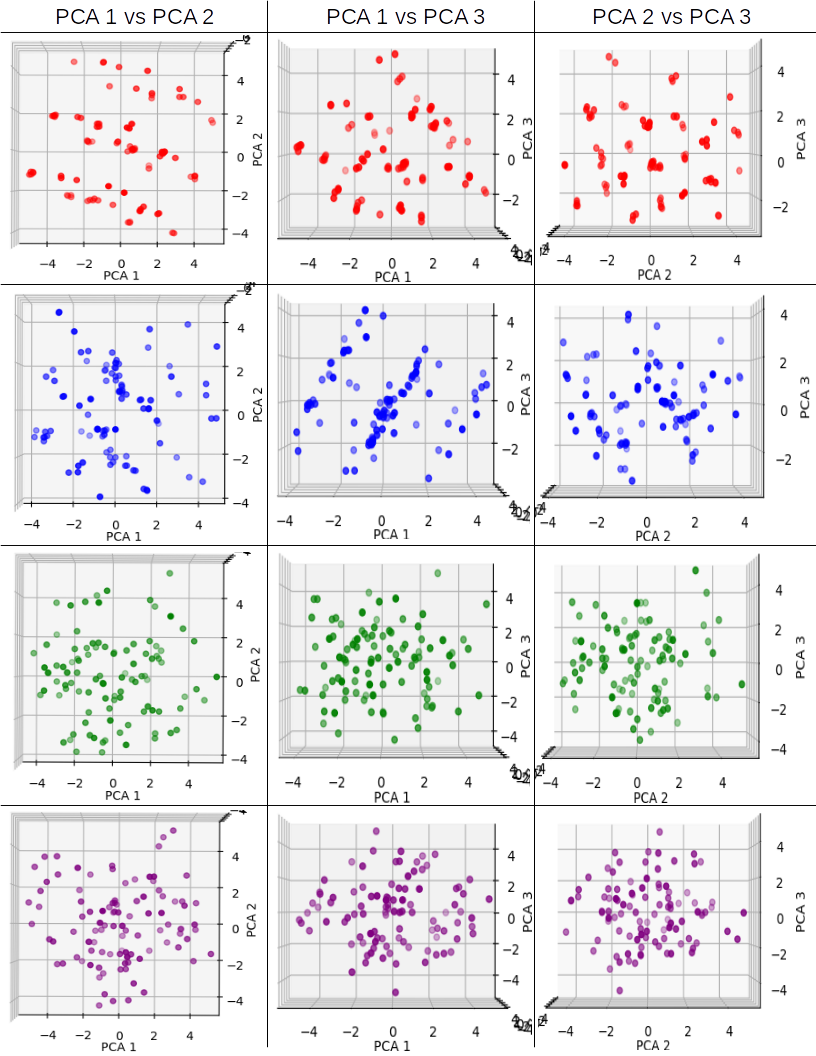}
\caption{PCA individual graphs for 15n complete configuration max-cut problem solved using QAOA, first 3 components. Red corresponds to the $3p$ parameter $1L$ non-entangled, blue $3p$ parameter $1L$ entangled, green $6p$ parameter $2L$ non-entangled and purple $6p$ parameter $2L$ entangled model.}
\label{fig:comp_ind_pca_15nCOM}
\end{figure}

The individual graphs using PCA for the 15n complete max-cut problem are presented in {Figure \ref{fig:comp_ind_pca_15nCOM}}. In the $3p$ models, both the entangled (blue) and non-entangled (red) models exhibit patterns similar to those observed in the previous 4n and 10n problems. However, there are some differences in the entangled model, particularly in the PCA 1 vs PCA 2 and PCA 1 vs PCA 3 planes, where more line patterns are observed compared to the one or two line patterns seen in the previous problems. Moving on to the $6p$ models, the non-entangled model (green) continues the trend observed in previous problems, showing no clear tendency or discernible behavior in the data distribution. In contrast, the entangled model (purple) exhibits no clear distribution in the PCA 1 vs PCA 2 plane, which is different from the patterns observed in the 4n and 10n problems. The PCA 1 vs PCA 3 plane shows some noisy cluster distribution, but it is not well-defined.

\begin{figure}[ht!]
\centering
\includegraphics[width=12cm, height=8cm]{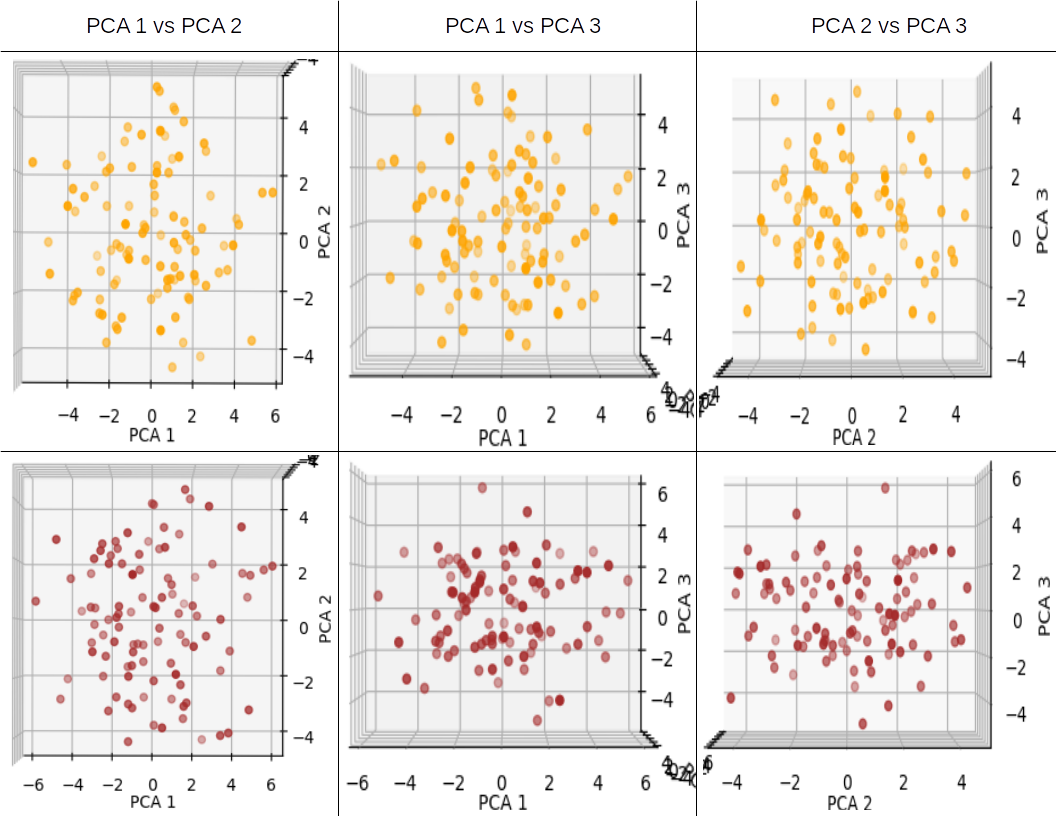}
\caption{PCA individual graphs for 15n complete configuration max-cut problem solved using QAOA, first 3 components. Orange $9p$ parameter $3L$ non-entangled and brown $9p$ parameter $3L$ entangled model.}
\label{fig:comp_ind_pca_15nCOM_2}
\end{figure}


In the $9p$ models depicted in {Figure \ref{fig:comp_ind_pca_15nCOM_2}}, no clear or distinguishable patterns can be observed in both the non-entangled (yellow) and entangled (brown) models. This behavior is consistent with the patterns observed in the 10n complete max-cut problem.

\begin{table}[ht!]
\centering
\begin{tabular}{|c|c|c|c|}
\hline
Parameters & PCA 1      & PCA 2      & PCA 3 \\ \hline
3 parameters         & 0.36035505 & 0.35400924 & 0.28563571 \\ \hline
6 parameters         & 0.24550126 & 0.19657842 & 0.16664964 \\ \hline
9 parameters         & 0.1512382 & 0.13967618 & 0.12791876 \\ \hline
\end{tabular}
\caption{Pair PCA projections explained variance for the first 3 PCA components for the 15n complete max-cut problem.}
\label{tab:exp_var_pairPCA_15nCOM}
\end{table}

The explained variance for the 15n complete max-cut problem is presented in Table \ref{tab:exp_var_pairPCA_15nCOM}. The distribution of PCA variance per model follows a similar trend as observed in the previous problems. The $3p$ models exhibit the highest variance values, which is expected as the number of parameters matches the number of PCA components. As the number of parameters increases, the quality of the components decreases, resulting in lower variance values. Notably, the $6p$ models show a slight increase in the total amount of variance compared to the 10n problem, bringing them closer to the values obtained in the cyclic problems. The variance values for the $9p$ models are similar to those observed in the 10n problems, both for cyclic and complete configurations, representing the lowest values among the tested models.

\begin{figure}[ht]
\centering
\includegraphics[width=10cm, height=8cm]{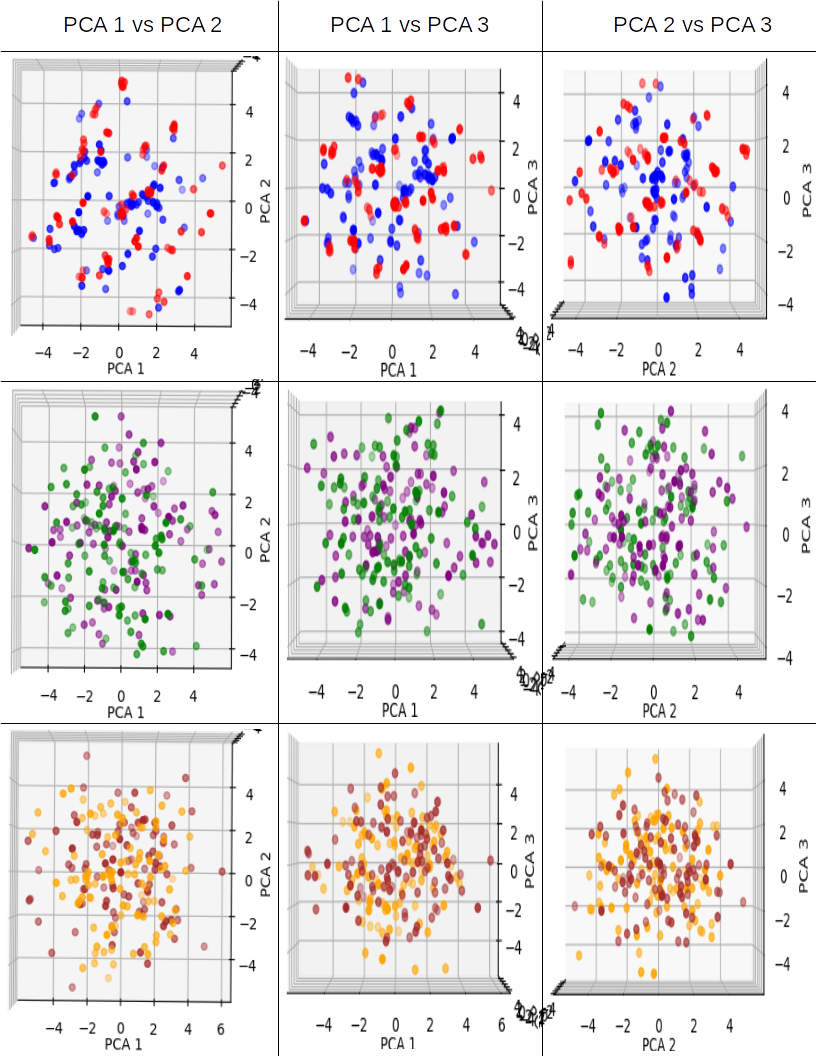}
\caption{PCA pair graphs for 15n complete configuration max-cut problem solved using QAOA, first 3 components. Red corresponds to the $3p$ parameter $1L$ non-entangled, blue $3p$ parameter $1L$ entangled, green $6p$ parameter $2L$ non-entangled, purple $6p$ parameter $2L$ entangled model, orange $9p$ parameter $3L$ non-entangled and brown $9p$ parameter $3L$ entangled model.}
\label{fig:comp_pair_pca_15n_COM}
\end{figure}

The pair PCA graphs for the 15n complete max-cut problem are presented in Figure \ref{fig:comp_pair_pca_15n_COM}. In the $3p$ models, both non-entangled (red) and entangled (blue), the essence of the individual graphs is preserved, similar to the previous pair graphs. However, the $6p$ models do not exhibit a clear behavior or pattern in any of the planes. This behavior is consistent with the 15n cyclic problem but differs from the distribution observed in the 4n and 10n complete problems, where some clustering patterns were observed. Lastly, in the $9p$ models, the non-entangled model (yellow) displays a random distribution pattern across all planes, while the entangled model (brown) shows a slightly more concentrated patterns in certain areas, as seen in the PCA 1 vs PCA 2 and PCA 1 vs PCA 3 planes.


\section{t-SNE graphs and KL divergence values} 

In this appendix, we present the complementary results for the experiments developed using t-SNE analysis (and KL-D values obtained) applied in the max-cut problems solved using QAOA.

\begin{table}[ht!]
\centering
\begin{tabular}{|c|c|c|c|}
\hline
Parameters & KL-D (3 per) & KL-D (30 per) & KL-D (99 per) \\ \hline
3 parameters         & 0.15324634   & 0.17391348    & 0.00002262    \\ \hline
3 parameters ent     & 0.14360289   & 0.05689293    & 0.00003242    \\ \hline
6 parameters         & 0.55960602   & 0.52287281     & 0.00004825    \\ \hline
6 parameters ent     & 0.37428555  & 0.38009176    & 0.00002795    \\ \hline
\end{tabular}
\caption{Individual KL-Divergence for 4n complete max-cut problem with different numbers of perplexity, considering the $3p$ non-entangled, $3p$ entangled, $6p$ non-entangled and $6p$ entangled models.}
\label{tab:ind_kl_divergence_t-SNE_4nCOM}
\end{table}

The results for KL-D for individual t-SNE in the 4n complete max-cut problem are presented in {Table \ref{tab:ind_kl_divergence_t-SNE_4nCOM}}. In general, the values for $3$ and $30$ perplexity have values that are closer compared to the cyclic 4n problem, with the entangled models having a better perplexity value (lower) compared to the non-entangled models. The best KL-D values were obtained with the $99$ perplexity, which is interpreted as the best model that represents the original properties of the data.

\begin{figure}[ht]
\centering
\includegraphics[width=10cm, height=8cm]{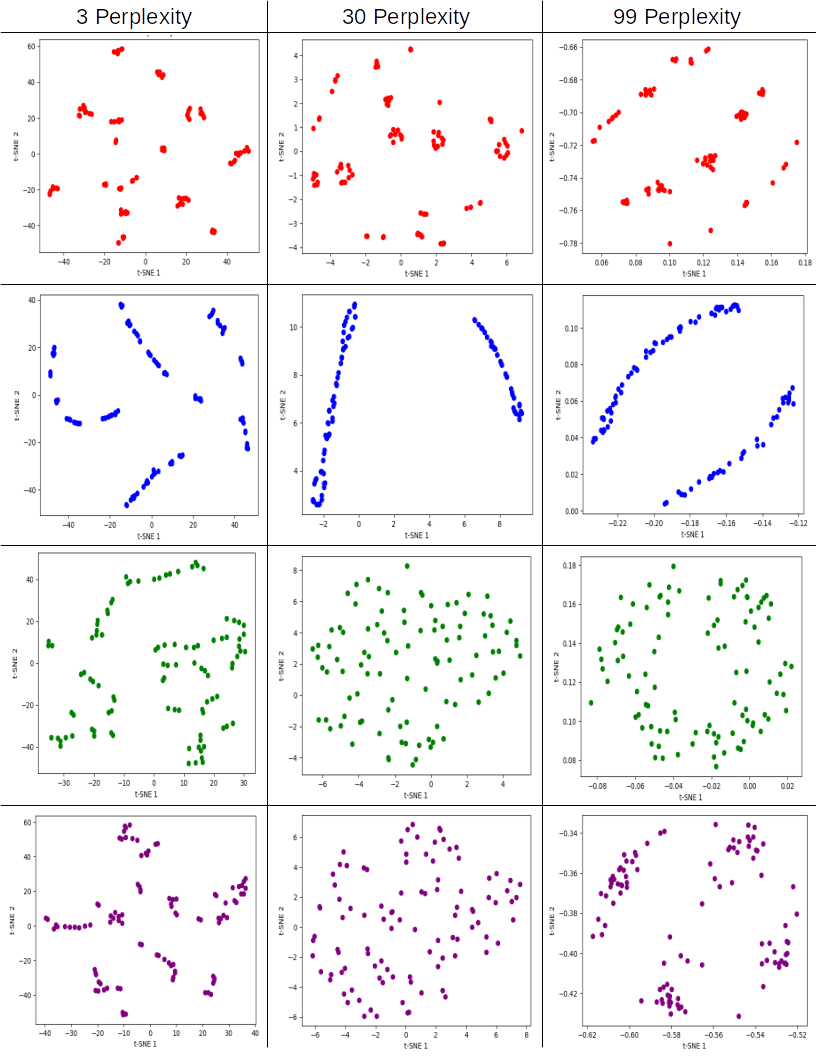}
\caption{t-SNE individual graphs for 4n complete configuration max-cut problem solved using QAOA, with different perplexity values $3$, $30$ and $99$. Red corresponds to the $3p$ parameter $1L$ non-entangled, blue $3p$ parameter $1L$ entangled, green $6p$ parameter $2L$ non-entangled and purple $6p$ parameter $2L$ entangled model.}
\label{fig:comp_ind_t-SNE_4nCOM}
\end{figure}

The individual graphs for the 4n complete max-cut problem ({Figure \ref{fig:comp_ind_t-SNE_4nCOM}}) show that the $3p$ non-entangled (red) model has a distribution similar to the previous problem, and specifically for the $99$ perplexity, the three-line pattern is similar to the one obtained before. The pattern in this perplexity value is also similar to some perspectives obtained in the PCA graphs. In the $3p$ entangled (blue) model has very different patterns than the ones obtained in the cyclic problem. The most interesting results are the similarities of the 2-line clusterization obtained in the $30$ and $99$ perplexity, which replicate some patterns from PCA graphs obtained in the same problem. Moving to the $6p$ models, the non-entangled (green) model has a random distribution behavior observed in the cyclic problem and the PCA graphs at the $30$ perplexity. At the $99$ perplexity, the elliptical pattern of the cyclic problem is observed again, but with a wider edge compared to the cyclic t-SNE graph. Last, for the $6p$ entangled (purple) model, the $99$ perplexity shows a particular pattern with two small elongated clusters at the extremes of the graph and two small clusters at the center of the plane with some outlier points trying to connect both small clusters.

\begin{table}[ht!]
\centering
\begin{tabular}{|c|cccc|}
\hline
             & \multicolumn{4}{c|}{KL-Divergence}                                                                               \\ \hline
Parameters   & \multicolumn{1}{c|}{3 per}      & \multicolumn{1}{c|}{30 per}     & \multicolumn{1}{c|}{99 per}     & 199 per    \\ \hline
3 parameters & \multicolumn{1}{c|}{0.17788552} & \multicolumn{1}{c|}{0.22446597} & \multicolumn{1}{c|}{0.11483472} & 0.00005328 \\ \hline
6 parameters & \multicolumn{1}{c|}{0.58645886} & \multicolumn{1}{c|}{0.77581}    & \multicolumn{1}{c|}{0.36295095} & 0.00005059 \\ \hline
\end{tabular}
\caption{Pair KL-Divergence for 4n complete max-cut problem with different numbers of perplexity, considering the $3p$ parameters (non-entangled and entangled) and $6p$ parameters (non-entangled and entangled) models.}
\label{tab:pair_kl_divergence_t-SNE_4nCOM}
\end{table}

The KL-D results for the pair t-SNE models in the 4n complete max-cut problem are presented in Table \ref{tab:pair_kl_divergence_t-SNE_4nCOM}. The results from $3$ to $99$ perplexity are quite similar to those obtained in the cyclic problem, where the $3p$ model shows better KL-D results, leading to a better representation of the data in the final plane. However, in the case of the $199$ perplexity, the $6p$ parameter models exhibit better KL-D values, which is a different result compared to the cyclic problem.

\begin{figure}[ht]
\centering
\includegraphics[width=10cm, height=8cm]{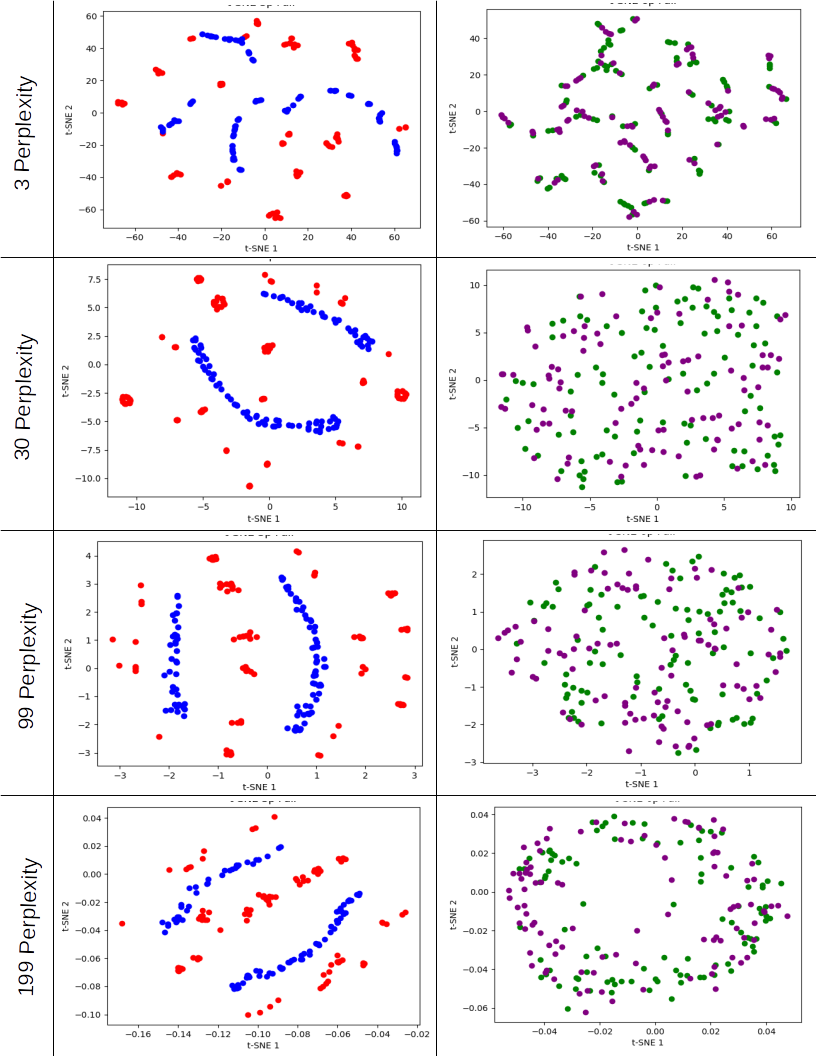}
\caption{t-SNE pair graphs for 4n pair complete configuration max-cut problem solved using QAOA, with different perplexity values $3$, $30$, $99$ and $199$. Red corresponds to the $3p$ parameter $1L$ non-entangled, blue $3p$ parameter $1L$ entangled, green $6p$ parameter $2L$ non-entangled and purple $6p$ parameter $2L$ entangled model.}
\label{fig:comp_pair_t-SNE_4nCOM}
\end{figure}

In the pair graphs for the 4n complete configuration ({Figure \ref{fig:comp_pair_t-SNE_4nCOM}}), we can observe interesting behavior patterns starting with the $3p$ models. The entangled model (blue) shows a similar pattern in all perplexity values, which can be observed more clearly from the $30$ to $199$ perplexity range. The blue model distributes itself over particular areas on the t-SNE mapped plane, but with a smooth distribution of mapped points. For the $6p$ models, the most interesting distribution is observed at $199$ perplexity value. Here, the mapped data distribution is very similar to the one obtained in the cyclic problem. However, in this case, there are only a few points that cross the middle of the elliptic pattern.

\begin{table}[ht!]
\centering
\begin{tabular}{|c|c|c|c|}
\hline
Parameters & KL-D (3 per) & KL-D (30 per) & KL-D (99 per) \\ \hline
3 parameters         & 0.11449474   & 0.18797217    & 0.00002232    \\ \hline
3 parameters ent     & 0.08714075   & 0.17103997    & 0.00004818    \\ \hline
6 parameters         & 0.61964202   & 0.53861362     & 0.00004456    \\ \hline
6 parameters ent     & 0.33782312  & 0.40035829    & 0.0000446    \\ \hline
9 parameters         & 0.6599322   & 0.60457009     & 0.00003925    \\ \hline
9 parameters ent     & 0.690759  & 0.54887885    & 0.00004668    \\ \hline
\end{tabular}
\caption{Individual KL-Divergence for 10n cyclic max-cut problem with different perplexity values, considering the $3p$ non-entangled, $3p$ entangled, $6p$ non-entangled, $6p$ entangled, $9p$ nonentangled and $9p$ entangled models.}
\label{tab:ind_kl_divergence_t-SNE_10nCYC}
\end{table}

When analyzing the 10n max-cut problem with cyclic configuration using t-SNE, we observed that the $3p$ entangled model performed the best in terms of KL-D value for the $3$ perplexity. However, as the number of parameters increased, the quality of the projected model decreased, but on average, entangled models produced better results than non-entangled ones. For the $30$ perplexity, the $3p$ entangled model remained the best, and all the entangled models had better KL-D results. At the $99$ perplexity, the $3p$ non-entangled model had the best KL-D value, but every model at this perplexity level showed a good KL-D value, which enabled a good representation of the data in the t-SNE plane.

\begin{figure}[ht]
\centering
\includegraphics[width=10cm, height=8cm]{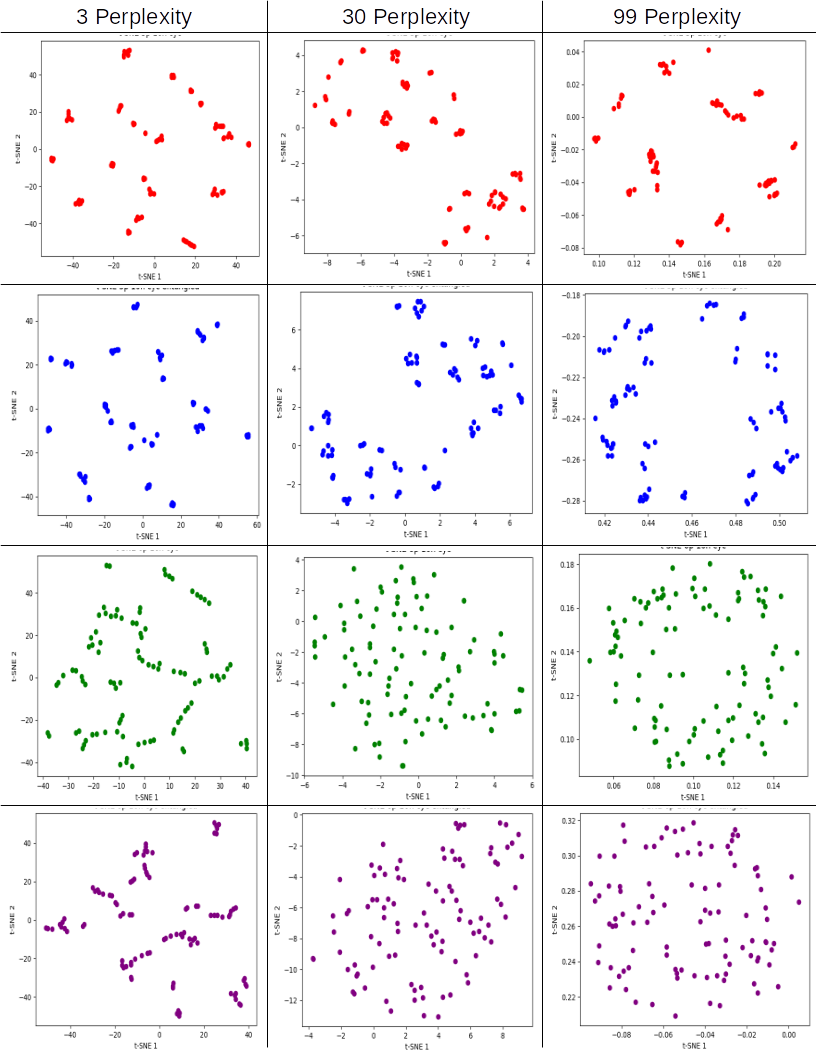}
\caption{t-SNE individual graphs for 10n cyclic configuration max-cut problem solved using QAOA, with different perplexity values $3$, $30$ and $99$. Red corresponds to the $3p$ parameter $1L$ non-entangled, blue $3p$ parameter $1L$ entangled, green $6p$ parameter $2L$ non-entangled and purple $6p$ parameter $2L$ entangled model.}
\label{fig:comp_ind_t-SNE_10nCYC}
\end{figure}

The graphs for the 10n cyclic max-cut problem are presented in {Figure \ref{fig:comp_ind_t-SNE_10nCYC}}. The patterns observed in the $3p$ models, both entangled (blue) and non-entangled (red), are pretty similar to the ones observed in the 4n problem. For the $6p$ non-entangled model (green), the distribution of data is similar to the one obtained in the 4n problem. However, for the entangled model (purple) at $99$ perplexity, the elliptic behavior is no longer distinguishable. In this case, the green and purple models at $99$ perplexity have a similar distribution of points.

\begin{figure}[ht]
\centering
\includegraphics[width=10cm, height=8cm]{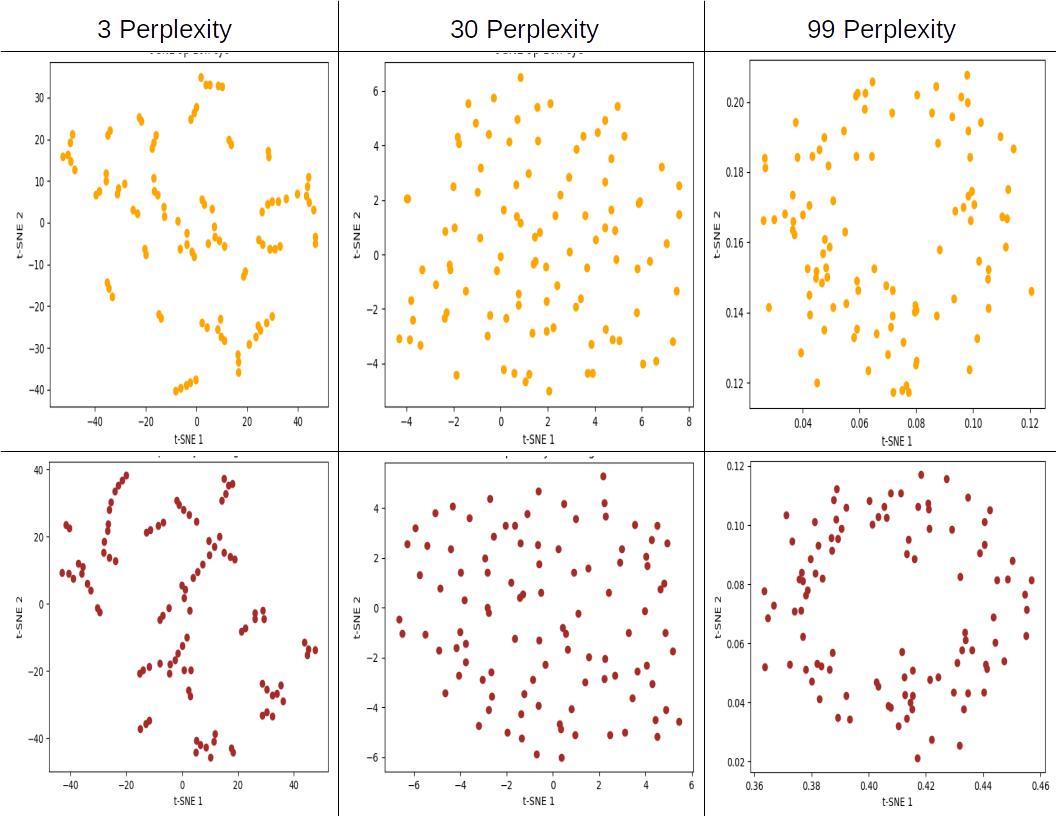}
\caption{t-SNE individual graphs for 10n cyclic configuration max-cut problem solved using QAOA, with different perplexity values $3$, $30$ and $99$. Orange corresponds to the $9p$ parameter $3L$ non-entangled and brown $9p$ parameter $3L$ entangled.}
\label{fig:comp_ind_t-SNE_10nCYC_2}
\end{figure}


In {Figure \ref{fig:comp_ind_t-SNE_10nCYC_2}}, at $99$ perplexity, the $9p$ non-entangled model (orange) exhibits a similar elliptic pattern as the $6p$ non-entangled model, while the entangled model (brown) displays a more defined elliptic pattern.

\begin{table}[ht!]
\centering
\begin{tabular}{|c|cccc|}
\hline
                                   & \multicolumn{4}{c|}{KL-Divergence}                                                                                                    \\ \hline
Parameters                         & \multicolumn{1}{c|}{3 per}      & \multicolumn{1}{c|}{30 per}     & \multicolumn{1}{c|}{99 per}     & 199 per                         \\ \hline
3 parameters                       & \multicolumn{1}{c|}{0.12718032} & \multicolumn{1}{c|}{0.22481607} & \multicolumn{1}{c|}{0.1436608}  & 0.00006457                      \\ \hline
6 parameters                       & \multicolumn{1}{c|}{0.61863911} & \multicolumn{1}{c|}{0.73845208} & \multicolumn{1}{c|}{0.37629709} & 0.0000484                       \\ \hline
\multicolumn{1}{|l|}{9 parameters} & \multicolumn{1}{l|}{0.80385178} & \multicolumn{1}{l|}{1.04350173} & \multicolumn{1}{l|}{0.423794}   & \multicolumn{1}{l|}{0.00004115} \\ \hline
\end{tabular}
\caption{Pair KL-Divergence for 10n cyclic max-cut problem with different perplexity values, considering the $3p$ parameters (non-entangled and entangled), $6p$ parameters (non-entangled and entangled) and $9p$ parameters (non-entangled and entangled) models.}
\label{tab:pair_kl_divergence_t-SNE_10nCYC}
\end{table}

{Table \ref{tab:pair_kl_divergence_t-SNE_10nCYC}} presents the pair KL-D divergences for different depth QAOA models. For the first three perplexity values ($3$, $30$, and $99$), the best KL-D values were obtained by the $3p$ models. At $30$ perplexity, it is interesting to see a value greater than 1 obtained by the $9p$ models, which is the highest (lower quality) value obtained so far. Finally, at $199$ perplexity, the best KL-D values for each t-SNE model were obtained, with the best KL-D value corresponding to the $9p$ models.

\begin{figure}[ht]
\centering
\includegraphics[width=10cm, height=8cm]{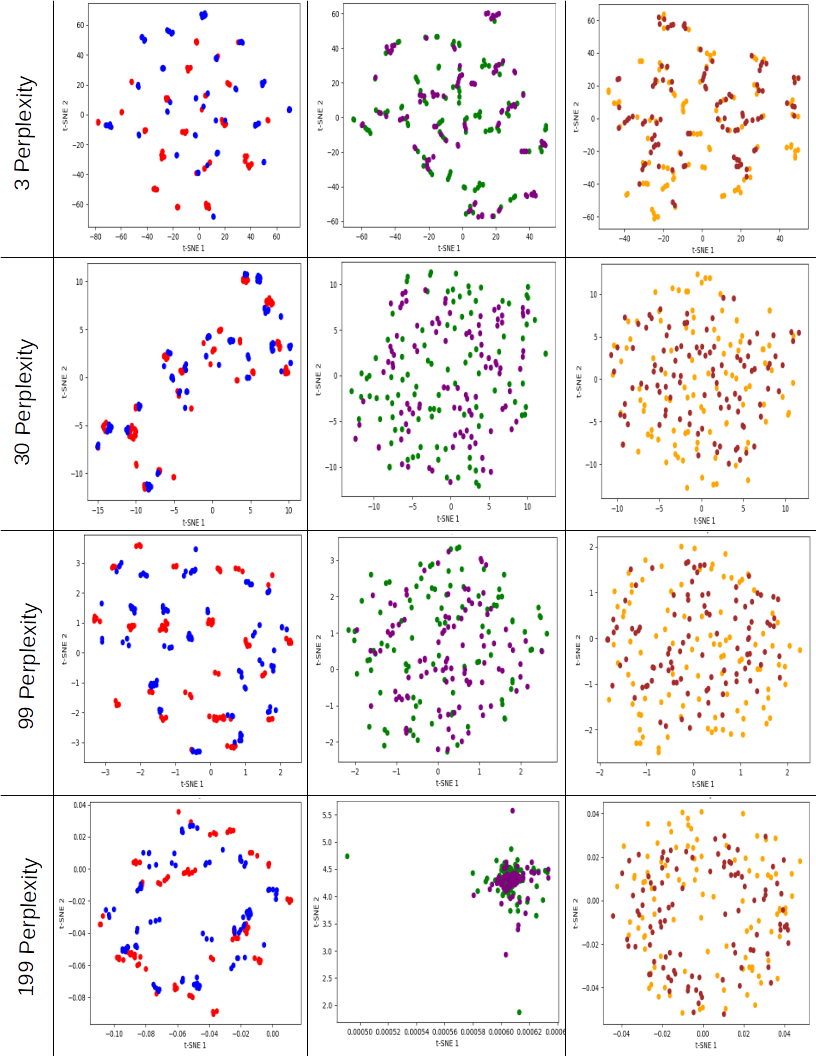}
\caption{t-SNE pair graphs for 10n cyclic configuration max-cut problem solved using QAOA, with different perplexity values $3$, $30$, $99$ and $199$. Red corresponds to the $3p$ parameter $1L$ non-entangled, blue $3p$ parameter $1L$ entangled, green $6p$ parameter $2L$ non-entangled, purple $6p$ parameter $2L$ entangled, orange $9p$ parameter $3L$ non-entangled and brown $9p$ parameter $3L$ entangled model.}
\label{fig:comp_pair_t-SNE_10nCYC}
\end{figure}

The pair t-SNE model graphs for the 10n cyclic max-cut problem can be seen in {Figure \ref{fig:comp_pair_t-SNE_10nCYC}}. In the $3$ perplexity, the non-entangled models (red, green, and orange) seem to be distributed in specific patterns in the plane, while the entangled models (blue, purple, and brown) match in certain areas of the t-SNE plane. Moving on to the $30$ perplexity, the $3p$ models (red non-entangled and blue entangled) distribute in very particular patterns that cannot be interpreted as a specific structure. In the $6p$ models, the green (non-entangled) model seems to have a smooth random distribution, while the purple (entangled) model is concentrated in certain areas of the plane. The $9p$ model follows a similar behavior as the $6p$ graph, where the orange (non-entangled) model is almost randomly distributed, and the brown (entangled) model is more concentrated. For the $99$ perplexity, the $3p$ graph has a similar pattern to the one observed in the 4n cyclic problem, where the red and blue models have fewer matches compared to the previous perplexity values. In the $6p$ graph, the behavior is similar to the one observed in the $30$ perplexity, where the green (non-entangled) model is scattered in the t-SNE plane, and the purple (entangled) model is more concentrated in certain areas. For the $9p$ graph, there is a difference between the orange (non-entangled) and brown (entangled) models, where the orange model maintains the scattered distribution, and the brown model has three areas where most of the points are plotted. Finally, in the $199$ perplexity, the $3p$ graph has a distribution that forms a rotated square with no additional specific behavior. The $6p$ graph has a completely different distribution from the ones observed in previous graphs, even in different problems. The scale of the graph is very small, generating the presence of outliers and a particular cluster containing both entangled (purple) and non-entangled (green) models. In the $9p$ graph, both orange (non-entangled) and brown (entangled) models have an elliptic pattern, where the orange model is more scattered compared to the brown model, which preserves the elliptic pattern better.

\begin{table}[ht!]
\centering
\begin{tabular}{|c|c|c|c|}
\hline
Parameters & KL-D (3 per) & KL-D (30 per) & KL-D (99 per) \\ \hline
3 parameters         & 0.1108679   & 0.15153457    & 0.00001909    \\ \hline
3 parameters ent     & 0.13275136   & 0.05032415    & 0.00004749    \\ \hline
6 parameters         & 0.48524341   & 0.51412958     & 0.00005262    \\ \hline
6 parameters ent     & 0.41168147  & 0.42243937    & 0.00003066    \\ \hline
9 parameters         & 0.73269081   & 0.6897254     & 0.00004309    \\ \hline
9 parameters ent     & 0.87109852  & 0.63736594    & 0.00004298    \\ \hline
\end{tabular}
\caption{Individual KL-Divergence for 10n complete max-cut problem with different perplexity values, considering the $3p$ non-entangled, $3p$ entangled, $6p$ non-entangled, $6p$ entangled, $9p$ non-entangled and $9p$ entangled models.}
\label{tab:ind_kl_divergence_t-SNE_10nCOM}
\end{table}


The KL-Divergence values presented in {Table \ref{tab:ind_kl_divergence_t-SNE_10nCOM}} show similar results to those observed in the 10n cyclic problem, where most of the entangled models present a better KL-Divergence value after optimization, resulting in a better mapping of points in the t-SNE plane.

\begin{figure}[ht]
\centering
\includegraphics[width=10cm, height=8cm]{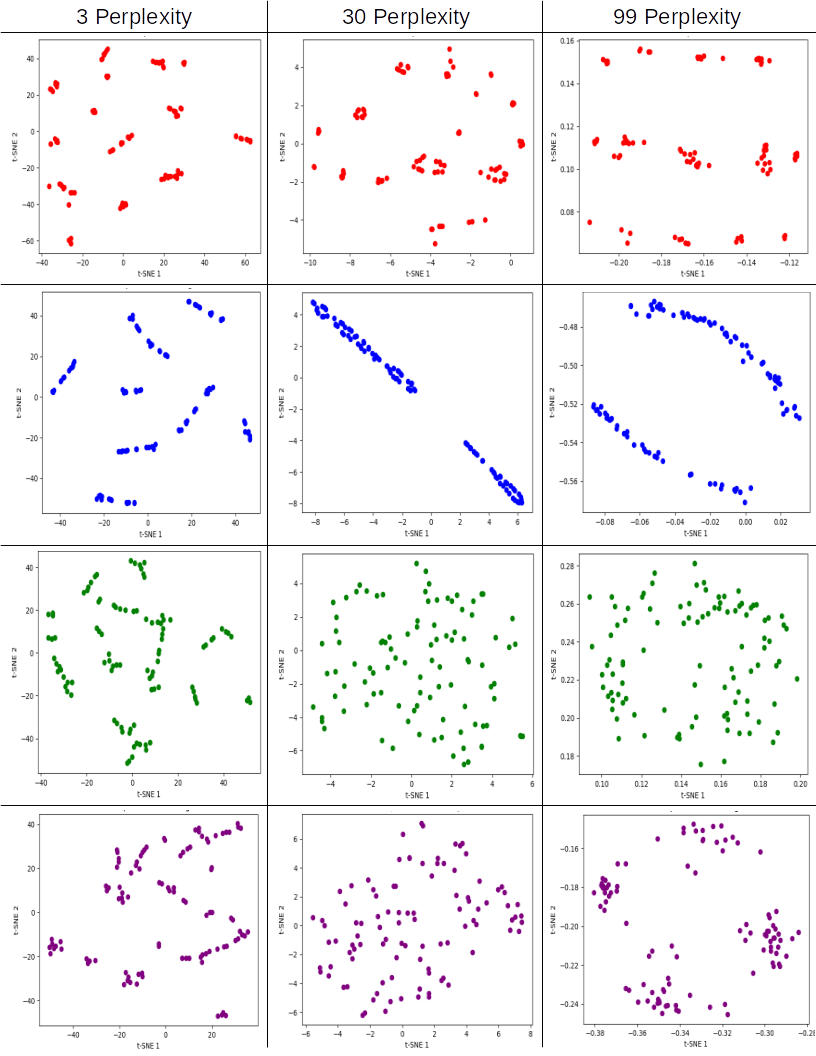}
\caption{t-SNE individual graphs for 10n complete configuration max-cut problem solved using QAOA, with different perplexity values $3$, $30$ and $99$. Red corresponds to the $3p$ parameter $1L$ non-entangled, blue $3p$ parameter $1L$ entangled, green $6p$ parameter $2L$ non-entangled and purple $6p$ parameter $2L$ entangled model.}
\label{fig:comp_ind_t-SNE_10nCOM}
\end{figure}

{Figure \ref{fig:comp_ind_t-SNE_10nCOM}} displays the individual t-SNE graphs for the 10n complete configuration max-cut problem. For the $3p$ models, the red (non-entangled) and blue (entangled) models at $3$ perplexity do not exhibit a clear pattern, consistent with previous results. At $30$ perplexity, the entangled model generates a line with an empty space in the middle, and the non-entangled model continues without a clear pattern. At $99$ perplexity, the non-entangled (red) model presents a pattern similar to the one seen in the 4n problem with complete configuration, as well as a similar pattern to the one obtained in the individual PCA graphs (PCA 1 vs PCA 2) for the 4n and 10n problems with a similar configuration. The entangled model (blue) at $99$ perplexity presents a two-line pattern, similar to the one obtained in the previous 4n problem and the individual PCA graphs (PCA 1 vs PCA 2) for the 4n and 10n complete configuration problems. For the $6p$ models, at $3$ and $30$ perplexity, there is no clear pattern, consistent with previous results. However, at $99$ perplexity, the non-entangled (green) model appears to be distributed in an elliptical pattern at the sides of the t-SNE plane, and the entangled (purple) model creates four clusters distributed at the sides of the plane. This last result shares some similarities with the 4n complete configuration problem.

\begin{figure}[ht]
\centering
\includegraphics[width=10cm, height=8cm]{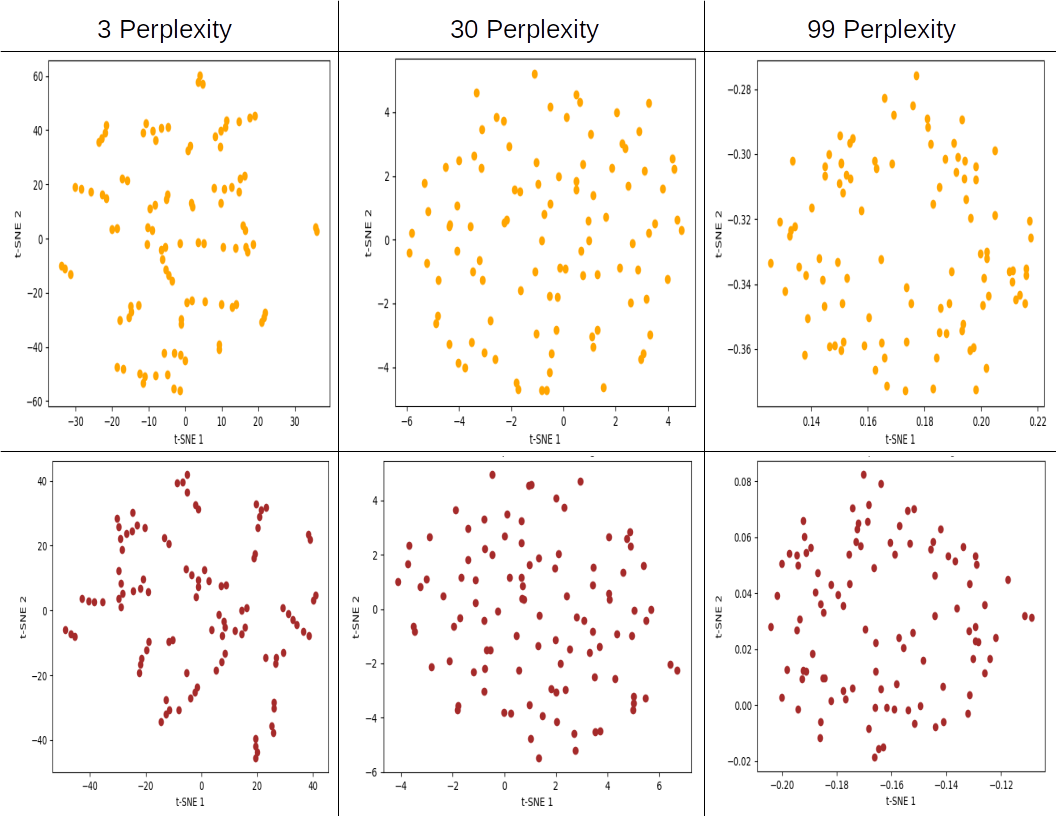}
\caption{t-SNE individual graphs for 10n complete configuration max-cut problem solved using QAOA, with different perplexity values $3$, $30$ and $99$. Orange corresponds to the $9p$ parameter $3L$ non-entangled and brown $9p$ parameter $3L$ entangled.}
\label{fig:comp_ind_t-SNE_10nCOM_2}
\end{figure}

The individual t-SNE graphs for the 10n complete configuration max-cut problem are shown in {Figure \ref{fig:comp_ind_pca_10nCOM_2}}. At both $3$ and $30$ perplexity, there is no clear pattern observed, with the distribution appearing random with no apparent clusters. At $99$ perplexity, there is also no distinguishable pattern observed, which is different from the elliptical behavior observed in the 10n cyclic problem but consistent with the individual PCA graphs obtained for the same problem.

\begin{table}[ht!]
\centering
\begin{tabular}{|c|cccc|}
\hline
                                   & \multicolumn{4}{c|}{KL-Divergence}                                                                                                    \\ \hline
Parameters                         & \multicolumn{1}{c|}{3 per}      & \multicolumn{1}{c|}{30 per}     & \multicolumn{1}{c|}{99 per}     & 199 per                         \\ \hline
3 parameters                       & \multicolumn{1}{c|}{0.15265435} & \multicolumn{1}{c|}{0.19996087} & \multicolumn{1}{c|}{0.11178039} & 0.00005902                      \\ \hline
6 parameters                       & \multicolumn{1}{c|}{0.56295419} & \multicolumn{1}{c|}{0.78792441} & \multicolumn{1}{c|}{0.38925377} & 0.00004848                      \\ \hline
\multicolumn{1}{|l|}{9 parameters} & \multicolumn{1}{l|}{0.90329468} & \multicolumn{1}{l|}{1.03177929} & \multicolumn{1}{l|}{0.43718094} & \multicolumn{1}{l|}{0.00004415} \\ \hline
\end{tabular}
\caption{Pair KL-Divergence for 10n complete max-cut problem with different perplexity values, considering the $3p$ parameters (non-entangled and entangled), $6p$ parameters (non-entangled and entangled) and $9p$ parameters (non-entangled and entangled) models.}
\label{tab:pair_kl_divergence_t-SNE_10nCOM}
\end{table}

The KL-Divergence values for the pair-wise t-SNE models are presented in {Table \ref{tab:pair_kl_divergence_t-SNE_10nCOM}}. The values are similar to those observed in the cyclic problem with 10n, where the worst KL-D values were obtained at 30 perplexity, particularly in the 9p parameter models, and the best KL-D values were obtained at 199 perplexity. The overall best performance was seen in the 9p models.

\begin{figure}[ht]
\centering
\includegraphics[width=10cm, height=8cm]{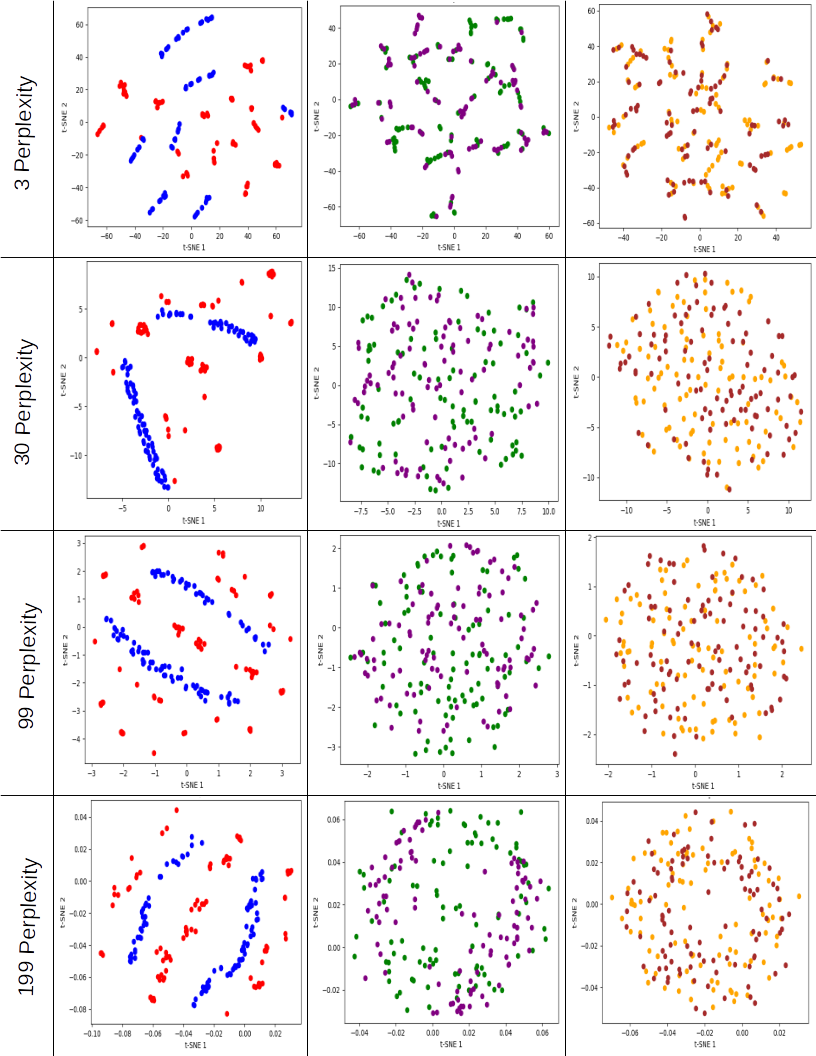}
\caption{t-SNE pair graphs for 10n complete configuration max-cut problem solved using QAOA, with different perplexity values $3$, $30$, $99$ and $199$. Red corresponds to the $3p$ parameter $1L$ non-entangled, blue $3p$ parameter $1L$ entangled, green $6p$ parameter $2L$ non-entangled, purple $6p$ parameter $2L$ entangled, orange $9p$ parameter $3L$ non-entangled and brown $9p$ parameter $3L$ entangled model.}
\label{fig:comp_pair_t-SNE_10nCOM}
\end{figure}

The pair graphs obtained for the 10n complete configuration problem can be seen in {Figure \ref{fig:comp_pair_pca_10n_COM}}. For the $3p$ non-entangled (red) and entangled (blue) models, similar patterns are observed as in the individual t-SNE graphs, where the entangled model preserves a two-line clusterization and the non-entangled model generates different types of lines that can be observed from $99$ to $199$ perplexity. It is also important to mention that the distribution for the $3p$ models is very similar to the ones observed in the 4n complete configuration problem in PCA 1 vs PCA 2 and the paired t-SNE graphs. For the $6p$ models, the most interesting behavior is presented at $199$ perplexity, where the non-entangled (green) model has an elliptical pattern with some points at the center, and the entangled (purple) model has two definite areas where the points are plotted, which are two parts of the elliptical pattern. This pattern has a lot of similarities with the 4n nodes problem at the same perplexity. For the $9p$ models, in general, the non-entangled model (orange) seems to be randomly distributed at different perplexities, where the entangled (brown) model tends to be more concentrated in certain areas of the t-SNE plane. At $199$ perplexity, both models tend to generate an elliptical behavior, where the non-entangled model is better distributed around the ellipse, and the entangled model is more scattered, this pattern is similar to the one observed in the $9p$ models for the 10n cyclic problem.


\begin{table}[ht!]
\centering
\begin{tabular}{|c|c|c|c|}
\hline
Parameters & KL-D (3 per) & KL-D (30 per) & KL-D (99 per) \\ \hline
3 parameters         & 0.12295903   & 0.19203556   & 0.0000241    \\ \hline
3 parameters ent     & 0.10832428   & 0.24514797   & 0.0000398    \\ \hline
6 parameters         & 0.63913888   & 0.5105567     & 0.00004167    \\ \hline
6 parameters ent     & 0.3930757  & 0.45114037    & 0.00003364    \\ \hline
9 parameters         & 0.71460283   & 0.66517001     & 0.00004309    \\ \hline
9 parameters ent     & 0.64783859  & 0.55502474    & 0.00004505    \\ \hline
\end{tabular}
\caption{Individual KL-Divergence for 15n cyclic max-cut problem with different perplexity values, considering the $3p$ non-entangled, $3p$ entangled, $6p$ non-entangled, $6p$ entangled, $9p$ non-entangled and $9p$ entangled models.}
\label{tab:ind_kl_divergence_t-SNE_15nCYC}
\end{table}

The KL-Divergence values presented in {Table \ref{tab:ind_kl_divergence_t-SNE_15nCYC}} correspond to the 15n cyclic problem. At a perplexity of $3$, the entangled approaches consistently produced better KL values across all models, with the best KL value obtained in the $3p$ entangled model. At a perplexity of $30$, the trend of entangled models performing better in terms of KL values continues for the more complex models with $6p$ and $9p$ ($2L$ and $3L$ depths, respectively). At a perplexity of $99$, all models exhibit good KL values, which are closer to zero. When comparing these results with those reported in the 10n cyclic problem, we observe a consistent trend where entangled models generally yield better KL-Divergence values for different perplexities. Additionally, the best KL values for mapping are obtained at a perplexity of $99$.

\begin{figure}[ht]
\centering
\includegraphics[width=10cm, height=8cm]{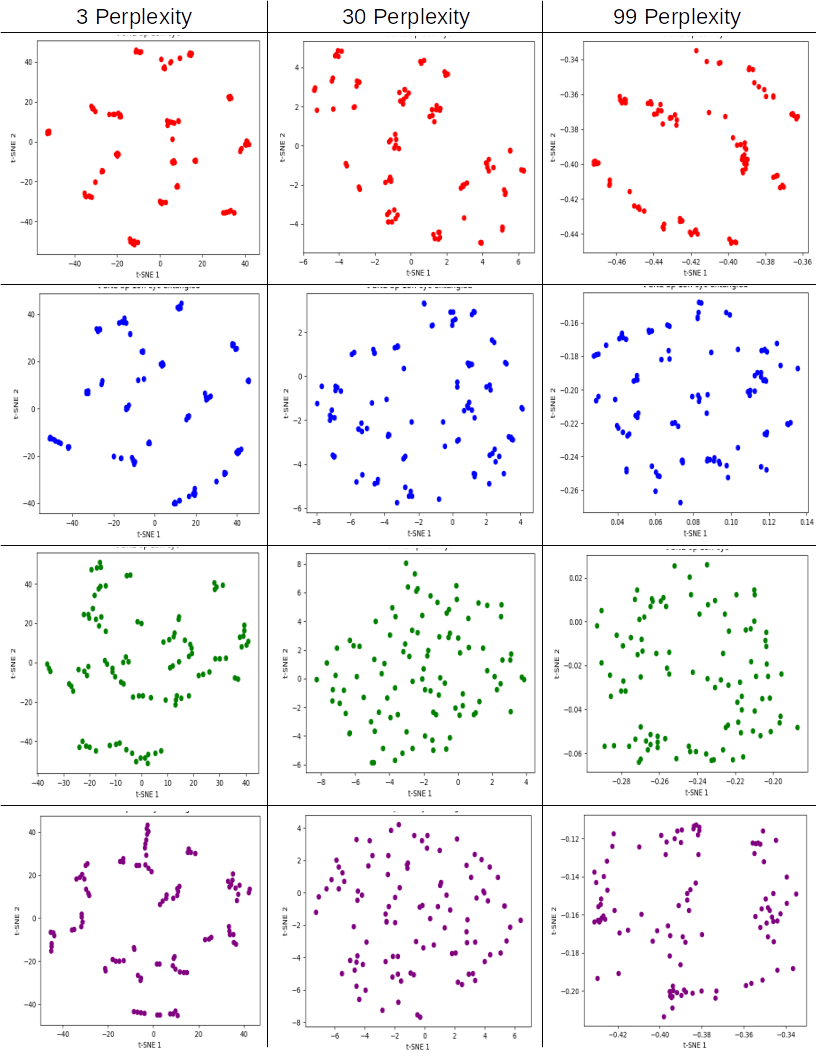}
\caption{t-SNE individual graphs for 15n cyclic configuration max-cut problem solved using QAOA, with different perplexity values $3$, $30$ and $99$. Red corresponds to the $3p$ parameter $1L$ non-entangled, blue $3p$ parameter $1L$ entangled, green $6p$ parameter $2L$ non-entangled and purple $6p$ parameter $2L$ entangled model.}
\label{fig:comp_ind_t-SNE_15nCYC}
\end{figure}


In the t-SNE individual graphs for the 15n cyclic max-cut problem ({Figure \ref{fig:comp_ind_t-SNE_15nCYC}}), we observe similar behaviors as in the previous 4n and 10n cyclic problems. For the $3p$ models, both the non-entangled (red) and entangled (blue) models exhibit different patterns at different perplexities, and at a perplexity of $99$, the non-entangled model generates the line pattern observed in previous t-SNE and PCA graphs. In the case of the $6p$ models, both the non-entangled (green) and entangled (purple) models show distributions that are consistent with previous problems. The non-entangled model generates an elliptic pattern with some points in the middle, while the entangled model exhibits a similar external pattern but with a more pronounced line in the middle.

\begin{figure}[ht]
\centering
\includegraphics[width=10cm, height=8cm]{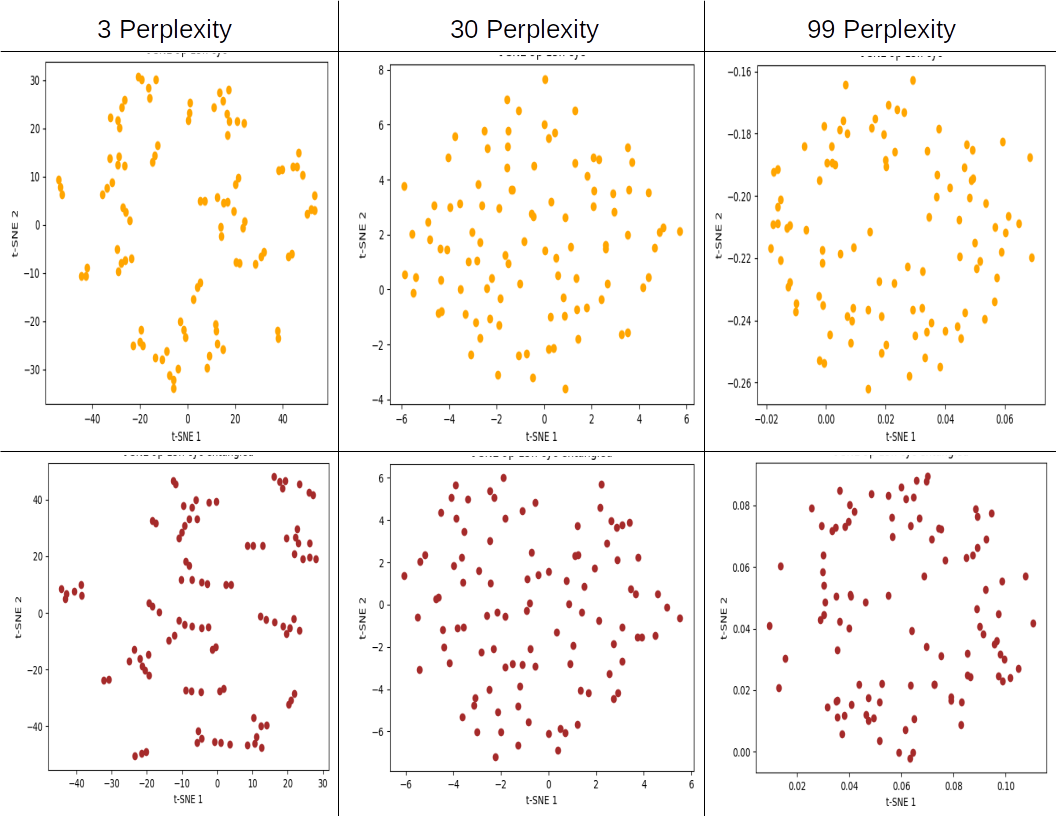}
\caption{t-SNE individual graphs for 15n cyclic configuration max-cut problem solved using QAOA, with different perplexity values $3$, $30$ and $99$. Orange corresponds to the $9p$ parameter $3L$ non-entangled and brown $9p$ parameter $3L$ entangled.}
\label{fig:comp_ind_t-SNE_15nCYC_2}
\end{figure}

The patterns observed in the $9p$ models at $3$ and $30$ perplexity ({Figure \ref{fig:comp_ind_t-SNE_15nCYC_2}}) closely resemble those observed in the 10n cyclic and complete configuration problems. At $99$ perplexity, the distribution of the non-entangled model (orange) is consistent with the previous problems, while the distribution of the entangled model (brown) follows a similar trend but with additional noise. The general pattern can still be perceived, but it is not as clear as in the previous cyclic problem.


\begin{table}[ht!]
\centering
\begin{tabular}{|c|cccc|}
\hline
                                   & \multicolumn{4}{c|}{KL-Divergence}                                                                                                    \\ \hline
Parameters                         & \multicolumn{1}{c|}{3 per}      & \multicolumn{1}{c|}{30 per}     & \multicolumn{1}{c|}{99 per}     & 199 per                         \\ \hline
3 parameters                       & \multicolumn{1}{c|}{0.13307488} & \multicolumn{1}{c|}{0.26993424} & \multicolumn{1}{c|}{0.17919816} & 0.00004499                      \\ \hline
6 parameters                       & \multicolumn{1}{c|}{0.64315343} & \multicolumn{1}{c|}{0.82122898} & \multicolumn{1}{c|}{0.34502453} & 0.00004286                      \\ \hline
\multicolumn{1}{|l|}{9 parameters} & \multicolumn{1}{l|}{0.8600843}  & \multicolumn{1}{l|}{1.03565741} & \multicolumn{1}{l|}{0.42468697} & \multicolumn{1}{l|}{0.00004554} \\ \hline
\end{tabular}
\caption{Pair KL-Divergence for 15n cyclic max-cut problem with different perplexity values, considering the $3p$ parameters (non-entangled and entangled), $6p$ parameters (non-entangled and entangled) and $9p$ parameters (non-entangled and entangled) models.}
\label{tab:pair_kl_divergence_t-SNE_15nCYC}
\end{table}

For the paired t-SNE models of the 15n cyclic max-cut problem presented in {Table \ref{tab:pair_kl_divergence_t-SNE_15nCYC}}, the observed values are similar to those of the 10n cyclic problem. At 3 perplexity, the best KL value corresponds to the paired 3p model, and as the number of parameters increases, the quality of KL-Divergence values decreases. At 30 perplexity, the best value is again obtained by the 3p models, but overall, this perplexity level yields the worst KL values. The trend of decreasing KL quality with an increasing number of parameters persists. At 99 perplexity, the values for the 6p and 9p models are improved compared to the previous perplexities, but the 3p models remain the best performers. Finally, at 199 perplexity, the overall best KL values are reported, with all models exhibiting good KL-Divergence values, indicating a well-mapped low-dimensional space where the 6p models yield the best KL value in this case.

\begin{figure}[ht]
\centering
\includegraphics[width=10cm, height=8cm]{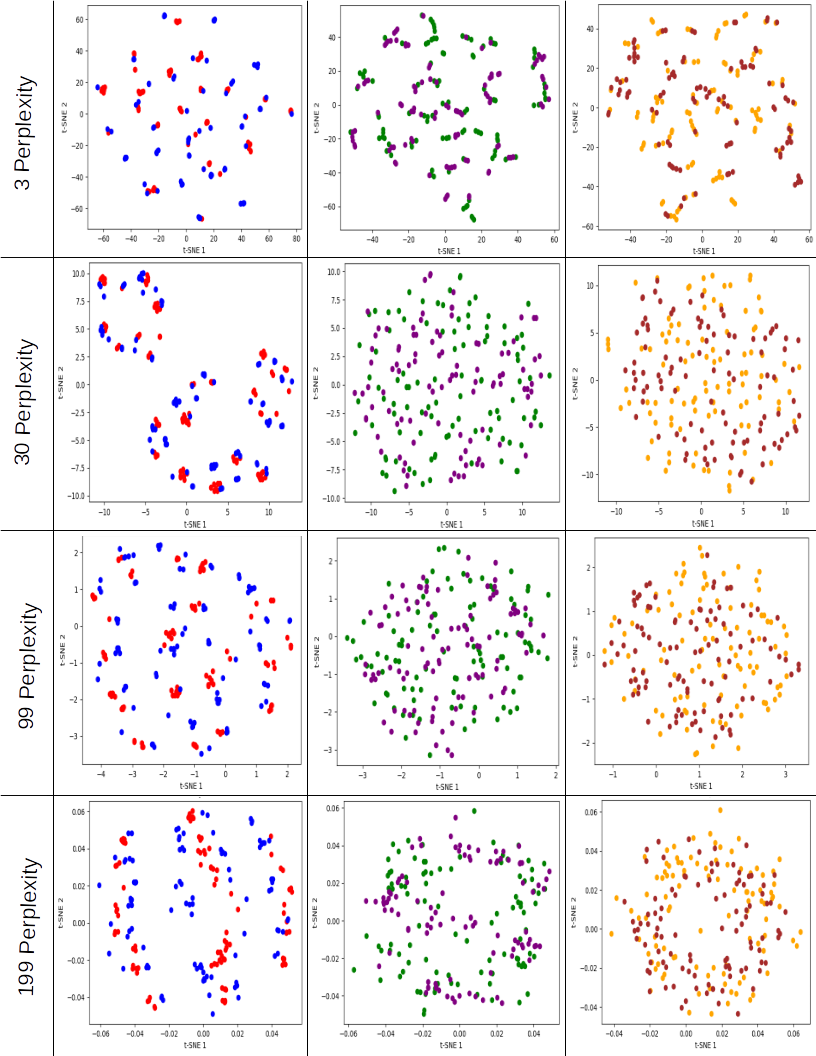}
\caption{t-SNE pair graphs for 15n cyclic configuration max-cut problem solved using QAOA, with different perplexity values $3$, $30$, $99$ and $199$. Red corresponds to the $3p$ parameter $1L$ non-entangled, blue $3p$ parameter $1L$ entangled, green $6p$ parameter $2L$ non-entangled, purple $6p$ parameter $2L$ entangled, orange $9p$ parameter $3L$ non-entangled and brown $9p$ parameter $3L$ entangled model.}
\label{fig:comp_pair_t-SNE_15nCYC}
\end{figure}

The graphical representation of the paired t-SNE models is presented in {Figure \ref{fig:comp_pair_t-SNE_15nCYC}}. Starting with the 3p models, both non-entangled (red) and entangled (blue), present similar behaviors as in the 10n cyclic case. From 3 to 99 perplexity values, the patterns of both models appear relatively similar, with each model tending to group more in certain areas. At 199 perplexity, the difference between models becomes more pronounced, where the non-entangled model exhibits a pattern with three lines, while the entangled model simulates a containment pattern of the non-entangled model. For the 6p models, the observed behaviors are also similar to those reported in the 10n cyclic problem. The non-entangled model (green) appears more scattered in the plane from 3 to 99 perplexity, while the entangled model (purple) tends to be more concentrated in certain areas. At 199 perplexity, the non-entangled and entangled models share a closer distribution, but the entangled model stands out due to the presence of three soft clusters. Finally, for the 9p models, the distributions are similar to the 6p models from 3 to 99 perplexity, where the non-entangled model (orange) shows a random distribution across most of the t-SNE plane, while the entangled model (brown) exhibits a higher concentration in certain areas. At 199 perplexity, both models generate an elliptical pattern, with the entangled model being more grouped in certain parts of the elliptical pattern.

\begin{table}[ht!]
\centering
\begin{tabular}{|c|c|c|c|}
\hline
Parameters & KL-D (3 per) & KL-D (30 per) & KL-D (99 per) \\ \hline
3 parameters         & 0.09598967   & 0.21283571   & 0.00005414    \\ \hline
3 parameters ent     & 0.21084341   & 0.16879296   & 0.00003727    \\ \hline
6 parameters         & 0.62666488   & 0.56735194     & 0.00003864    \\ \hline
6 parameters ent     & 0.34622833  & 0.42480648    & 0.00004639    \\ \hline
9 parameters         & 0.82630664   & 0.66207534     & 0.00004564    \\ \hline
9 parameters ent     & 0.66798007  & 0.57983494    & 0.00004709    \\ \hline
\end{tabular}
\caption{Individual KL-Divergence for 15n complete max-cut problem with different perplexity values, considering the $3p$ non-entangled, $3p$ entangled, $6p$ non-entangled, $6p$ entangled, $9p$ non-entangled and $9p$ entangled models.}
\label{tab:ind_kl_divergence_t-SNE_15nCOM}
\end{table}

The last individual t-SNE KL-Divergence values correspond to the 15n complete max-cut problem, and they are presented in {Table \ref{tab:ind_kl_divergence_t-SNE_15nCOM}}. At 3 perplexity, the entangled models for $6p$ and $9p$ present better KL values. However, for the $3p$ models, the non-entangled model has the best KL value overall, which differs from the values observed in the 10n complete problem where only the $6p$ entangled model was better than the non-entangled model. At 30 perplexity, all the entangled models show better values compared to their corresponding non-entangled models. This behavior is similar to what was observed in the 10n complete problem at the same perplexity. Finally, at 99 perplexity, all the models exhibit good KL-Divergence values, with the best value obtained by the $3p$ entangled model. Overall, the KL values for this problem demonstrate better results for the entangled models. They also share more similarities with the values observed in the 15n cyclic problem and, at certain perplexities, with the 10n complete problem.

\begin{figure}[ht]
\centering
\includegraphics[width=10cm, height=8cm]{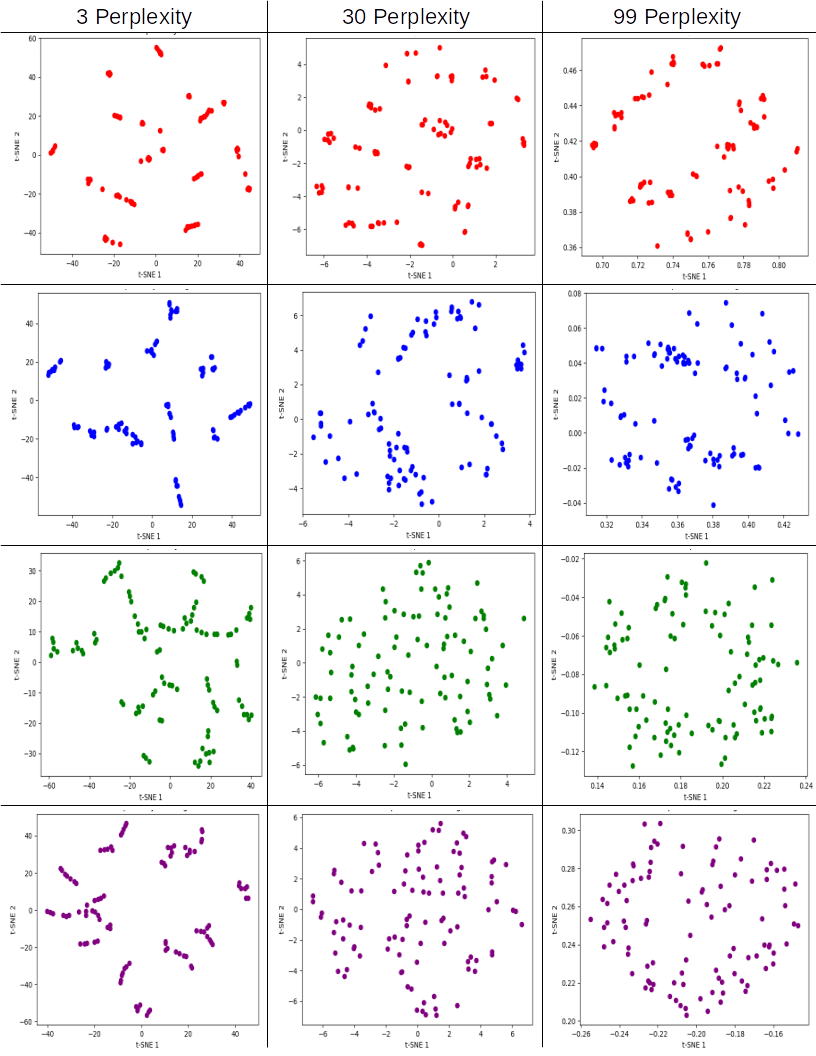}
\caption{t-SNE individual graphs for 15n complete configuration max-cut problem solved using QAOA, with different perplexity values $3$, $30$ and $99$. Red corresponds to the $3p$ parameter $1L$ non-entangled, blue $3p$ parameter $1L$ entangled, green $6p$ parameter $2L$ non-entangled and purple $6p$ parameter $2L$ entangled model.}
\label{fig:comp_ind_t-SNE_15nCOM}
\end{figure}

The graphs for the 15n complete max-cut problem can be viewed in {Figure \ref{fig:comp_ind_t-SNE_15nCOM}}. For the $3p$ models, non-entangled (red) and entangled (blue), at 3 perplexity, we observe similar patterns as those observed in previous problems. At 30 perplexity, the distribution is different from what was observed in the 10n complete problem, resembling the pattern observed in the 15n cyclic problem. At 99 perplexity, the non-entangled model exhibits a similar 3 line pattern as in previous problems, but the entangled model shows a distribution with two separate areas from the middle, forming line patterns. For the $6p$ models, non-entangled (green) and entangled (purple), the behavior at 3 and 30 perplexity is similar to what was reported in the 10n complete and 15n cyclic problems. At 99 perplexity, the non-entangled model displays an elliptic pattern with some random points around it, while the entangled model generates a deformed elliptic pattern, resembling a butterfly-like distribution.

\begin{figure}[ht]
\centering
\includegraphics[width=10cm, height=8cm]{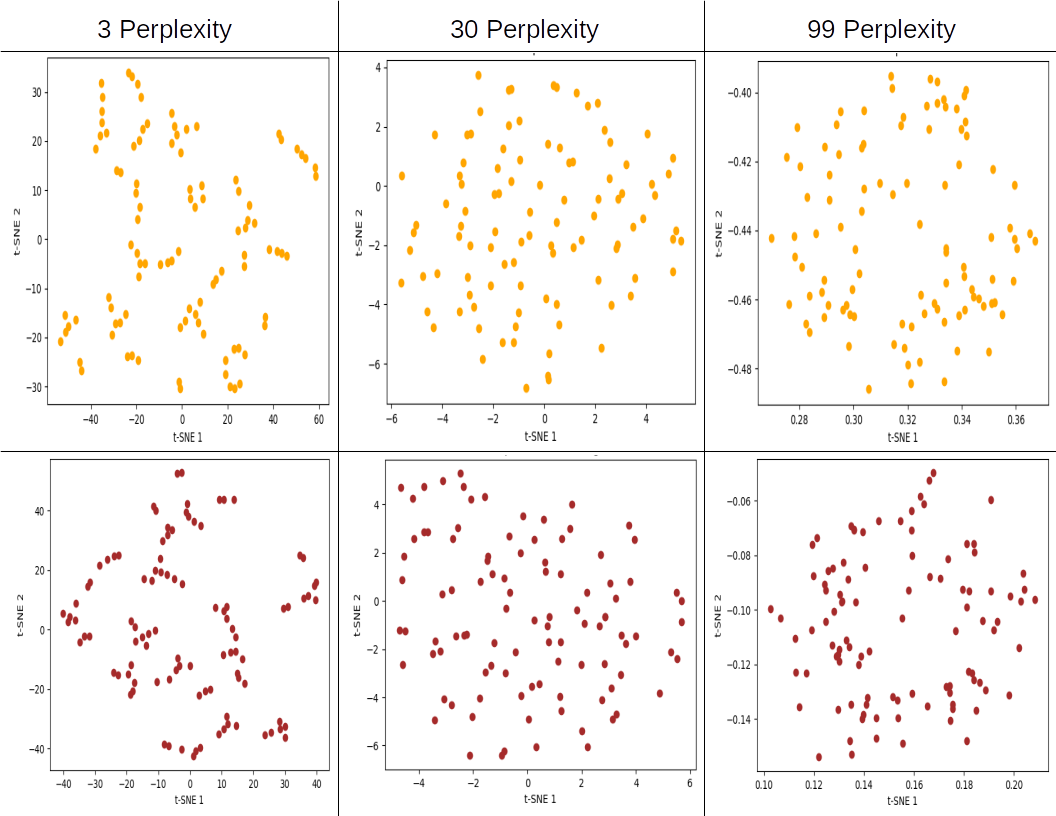}
\caption{t-SNE individual graphs for 15n complete configuration max-cut problem solved using QAOA, with different perplexity values $3$, $30$ and $99$. Orange corresponds to the $9p$ parameter $3L$ non-entangled and brown $9p$ parameter $3L$ entangled.}
\label{fig:comp_ind_t-SNE_15nCOM_2}
\end{figure}

Finally, for the $9p$ models, the graphical results are presented in {Figure \ref{fig:comp_ind_t-SNE_15nCOM_2}}. At 3 and 30 perplexity, the patterns observed for the non-entangled (orange) and entangled (brown) models are similar to the ones observed in the 10n complete and 15n cyclic problems. At 99 perplexity, both the non-entangled and entangled models exhibit a tendency to concentrate more towards the sides of the t-SNE plane, creating a somewhat elliptical pattern that is not very distinct.

\begin{table}[ht!]
\centering
\begin{tabular}{|c|cccc|}
\hline
             & \multicolumn{4}{c|}{KL-Divergence}                                                                               \\ \hline
Parameters   & \multicolumn{1}{c|}{3 per}      & \multicolumn{1}{c|}{30 per}     & \multicolumn{1}{c|}{99 per}     & 199 per    \\ \hline
3 parameters & \multicolumn{1}{c|}{0.15160248} & \multicolumn{1}{c|}{0.26059961} & \multicolumn{1}{c|}{0.165535}   & 0.00004839 \\ \hline
6 parameters & \multicolumn{1}{c|}{0.6180442}  & \multicolumn{1}{c|}{0.75135112} & \multicolumn{1}{c|}{0.34869462} & 0.0000503  \\ \hline
9 parameters & \multicolumn{1}{c|}{0.943533}   & \multicolumn{1}{c|}{1.02061999} & \multicolumn{1}{c|}{0.43895942} & 0.00003921 \\ \hline
\end{tabular}
\caption{Pair KL-Divergence for 15n complete max-cut problem with different perplexity values, considering the $3p$ parameters (non-entangled and entangled), $6p$ parameters (non-entangled and entangled) and $9p$ parameters (non-entangled and entangled) models.}
\label{tab:pair_kl_divergence_t-SNE_15nCOM}
\end{table}

The KL-Divergence values for the paired t-SNE models in the 15n complete max-cut problem are presented in {Table \ref{tab:pair_kl_divergence_t-SNE_15nCOM}}. The values at 3, 30, and 99 perplexity exhibit similar behaviors as the 15n cyclic problem, where the best KL value was generated by the $3p$ models and the worst values were obtained at 30 perplexity for the $9p$ models specifically. Furthermore, at 199 perplexity, the best KL values were reported, with all the models generating good values and the best among them being the 9p models.

\begin{figure}[ht]
\centering
\includegraphics[width=10cm, height=8cm]{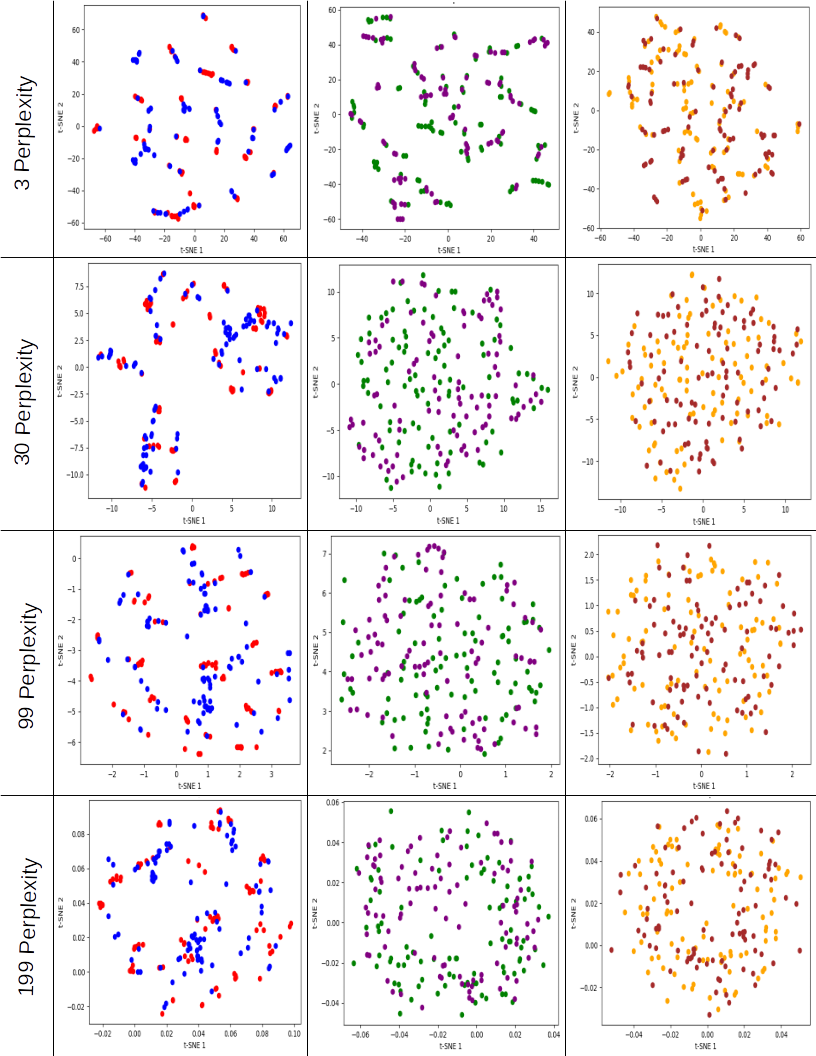}
\caption{t-SNE pair graphs for 15n complete configuration max-cut problem solved using QAOA, with different perplexity values $3$, $30$, $99$ and $199$. Red corresponds to the $3p$ parameter $1L$ non-entangled, blue $3p$ parameter $1L$ entangled, green $6p$ parameter $2L$ non-entangled, purple $6p$ parameter $2L$ entangled, orange $9p$ parameter $3L$ non-entangled and brown $9p$ parameter $3L$ entangled model.}
\label{fig:comp_pair_t-SNE_15nCOM}
\end{figure}

The paired t-SNE models for the 15n complete max-cut problem are presented in {Figure \ref{fig:comp_pair_t-SNE_15nCOM}}. In the $3p$ models, non-entangled (red) and entangled (blue), the behavior observed at different perplexities is very similar between them, with no clear distribution even at 199 perplexity. This result differs from the patterns observed in the 10n complete problem and the 15n cyclic problem. Moving on to the $6p$ models, the patterns observed in the non-entangled (green) and entangled (purple) models are consistent with the previous graphs. The non-entangled model tends to be randomly scattered across the plane, while the entangled model shows more grouping behavior at 3, 30, and 99 perplexities. Only at 199 perplexity do the models distribute themselves at the sides of the plane, with the entangled model being more concentrated in certain areas of the distribution. Finally, for the $9p$ models, at 3, 30, and 99 perplexities, the non-entangled (orange) and entangled (brown) models exhibit similar distributions to the $6p$ models. The non-entangled model is more scattered, while the entangled model generates small group patterns in certain areas of the plane. At 199 perplexity, both models exhibit some sort of elliptical pattern previously observed in other problems, with the non-entangled model showing a more pronounced elliptic shape and the entangled model following the pattern but with less smoothness.

\end{document}